\documentclass[useAMS,usenatbib,usegraphicx]{mn2e}
\usepackage{color}

\newcommand\beq{\begin{equation}}
\newcommand\eeq{\end{equation}}

\newcommand{\be}{\begin{equation}}
\newcommand{\ee}{\end{equation}}
\newcommand{\ba}{\begin{eqnarray}}
\newcommand{\ea}{\end{eqnarray}}
\newcommand{\Mc}{{\cal M}}
\newcommand{\Ms}{M_{\odot}}

\newcommand{\bml}{\begin{mathletters}}
\newcommand{\eml}{\end{mathletters}}

\newcommand{\Gpc}{\,\mathrm{Gpc}}
\newcommand{\yr}{\,\mathrm{yr}}

\newcommand{\pc}{\,\mathrm{pc}}
\newcommand{\Hz}{\,\mathrm{Hz}}

\newcommand{\D}{\mathrm{d}}
\newcommand{\tres}{t_{\rm res}}
\newcommand{\torb}{t_{\rm orb}}

\newcommand{\gw}{{\rm gw}}

\def\ltsima{$\; \buildrel < \over \sim \;$}
\def\simlt{\lower.5ex\hbox{\ltsima}}
\def\gtsima{$\; \buildrel > \over \sim \;$}
\def\simgt{\lower.5ex\hbox{\gtsima}}
\def\gsim{ \lower .75ex \hbox{$\sim$} \llap{\raise .27ex \hbox{$>$}} }
\def\lsim{ \lower .75ex\hbox{$\sim$} \llap{\raise .27ex \hbox{$<$}} }
\def\msun{\,{\rm M_\odot}}

\title[Massive black holes and pulsar timing]
{Gas driven massive black hole binaries: signatures in the nHz gravitational wave background}
\author[Bence Kocsis \& Alberto Sesana]{
B. Kocsis$^{1,2,3,4}$\thanks{E-mail: bkocsis@cfa.harvard.edu}
and A. Sesana$^{5,6}$\thanks{E-mail: alberto.sesana@aei.mpg.de}\\
$^{1}$Harvard-Smithsonian Center for Astrophysics, 60 Garden Street,
Cambridge, MA 02138, USA\\
$^{2}$Institute for Advanced Study, Einstein Dr,
Princeton, NJ, 08540, USA\\
$^{3}$E\"otv\"os University, P\'azm\'any P. st. 1/A, Budapest H-1117, Hungary\\
$^{4}$Einstein Fellow
\\
$^{5}$Albert Einstein Institute, Am Muhlenberg 1 D-14476 Golm, Germany\\
$^{6}$Center for Gravitational Wave Physics, The Pennsylvania State University, University Park, PA 16802, USA}

\begin{document}

\date{Received ---}

\maketitle
\begin{abstract}
Pulsar timing arrays (PTAs) measure nHz frequency gravitational waves (GWs) generated by orbiting massive
black hole binaries (MBHBs) with periods between 0.1 -- 10 yr. Previous studies on the nHz GW background assumed
that the inspiral is purely driven by GWs. However, torques generated by a gaseous disk can shrink the binary much more
efficiently than GW emission, reducing the number of binaries at these
separations. We use simple disk models for the circumbinary gas and for the binary-disk interaction to follow the orbital decay of
MBHBs through physically distinct regions of the disk, until GWs take over their evolution. We extract MBHB
cosmological merger rates from the Millennium simulation, generate Monte Carlo realizations of a population of gas driven
binaries, and calculate the corresponding GW amplitudes of the most luminous individual binaries and the
stochastic GW background. For steady--state $\alpha$--disks with $\alpha>0.1$ we find that the nHz GW background can be significantly
modified. The number of resolvable binaries is however not changed by the presence of gas; we predict 1--10 individually resolvable
sources to stand above the noise for a 1--50\,ns timing precision. Gas driven migration
reduces predominantly the number of small total mass or unequal mass ratio binaries, which leads to the attenuation
of the mean stochastic GW--background, but increases the detection significance of individually resolvable binaries.
{ The results are sensitive to the model of binary--disk interaction. The GW background is not attenuated significantly for  
time-dependent models of \citet{ivanov}.}
\end{abstract}

\section{Introduction}\label{s:introduction}
Inspiraling massive black hole binaries (MBHBs) with masses in the range $\sim 10^4-10^{10}\msun$ are
expected to be the dominant source of gravitational waves (GWs) at $\sim$ nHz -- mHz frequencies
\citep{Haehnelt94,Jaffe03,Wyithe03,Sesana04,Sesana05}.
The frequency band $\sim 10^{-5}\,\mathrm{Hz}$ -- $1 \,\mathrm{Hz}$
will be probed by the {\it Laser Interferometer Space Antenna} ({\it LISA}, \citealt{Bender98}), a
space-borne GW laser interferometer developed by ESA and NASA. The observational window
$10^{-9}\,\mathrm{Hz}$ -- $10^{-6} \,\mathrm{Hz}$, corresponding roughly to orbital periods 0.03 -- $30\yr$,
is already accessible using Pulsar Timing Arrays
(PTAs;  e.g. the Parkes radio-telescope, \citealt{Manchester08}).
The complete Parkes PTA  (PPTA, \citealt{Manchester08}), the European Pulsar Timing Array (EPTA, \citealt{Janssen08}), and NanoGrav
\citep{jen09} are expected to improve considerably
on the capabilities of these surveys, eventually joining their efforts in the
international PTA project (IPTA, \citealt{Hobbs09});
and the planned Square Kilometer Array (SKA, \citealt{Lazio09})
will produce a major leap in sensitivity.

Radio pulses generated by rotating neutron stars travel through the Galactic interstellar medium and are detected
by radio telescopes on Earth. The arrival times of pulses are fitted for a model including all the known and
measured systematic effects affecting the signal generation, propagation and detection \citep{ed06}.
Timing residuals between the observed
pulses and the best fit model, carry information on additional unmodelled effects, including the presence of GWs.
Indeed, GWs modify the propagation of radio signals from the pulsar
to the Earth \citep{Sazhin78,Detweiler79,Bertotti83,Hellings83,Jenet05},
and PTAs measure the direction dependent systematic variations in the arrival times of signals from
a sample of nearly stationary pulsars in the Galaxy distributed over the sky.

PTAs provide a direct observational window onto the MBH binary population,
and can contribute to address a number of open astrophysical questions, such as the shape
of the bright end of the MBH mass function, the nature of the MBH-bulge relation
at high masses, and the dynamical evolution at sub-parsec scales of the most
massive binaries in the Universe  (particularly relevant to the so-called ``final parsec problem'',
\citealt{Milos03})
PTAs can detect gravitational radiation of two forms: (i) the stochastic GW background produced by the
incoherent superposition of radiation from the whole cosmic population of MBHBs and
(ii) individual sources that are sufficiently bright in GWs to outshine the background
(typically massive, $M\gsim 10^{9}\msun$, and ``cosmologically nearby'', $d_L \lsim 3\Gpc$).
Both classes of signals are of great interest, and PTAs could lead to the discovery of systems
difficult to detect with other techniques
\citep[for alternatives using active galactic nuclei, see][and references therein]{HKM09}.

Popular scenarios of massive black hole (MBH) formation and evolution
\citep[e.g.][]{VHM03,Koushiappas06,Malbon07,Yoo07} predict frequent MBH mergers (up to
several hundreds per year), implying the existence of a large number of sub-parsec MBHBs.
The prospect for detecting GW signals using PTAs depends on the number and cosmological distribution
of MBHBs with orbital periods of 0.03 -- $30\yr$, or separations typically in the range $0.001-0.1\pc$.
The three main ingredients for calculating the GW background are
\begin{enumerate}
\item[(i)] the merger rate of MBHBs as a function of mass and redshift,
\item[(ii)] the relative time each binary spends at these separations during a merger episode
and
\item[(iii)] the amplitude of the GW signal produced by each individual stationary system.
\end{enumerate}

Recently \citet[SVC08 hereinafter]{SVC08} and \citet[SVV09 hereinafter]{SVV09},
carried out a detailed study of the expected signals
(stochastic and individual), focusing on the uncertainties related to (i).
They found that the background is affected by the
galaxy merger rate evolution along the cosmic history, the massive black hole mass function,
and the accretion history of the MBHB during a galaxy merger, and they predict a factor of
$\sim 10$ uncertainty for the characteristic strain amplitude in the range
$2\times 10^{-15}-2\times10^{-16}$, at $f=1/$yr, within the expected detection capabilities of the
complete PPTA and of the SKA. They pointed out that the GW signal can be separated into individually resolvable sources
and a stochastic background, and found the number of individually resolvable sources for a 1ns timing
precision level to be between 5 to 15, depending on the considered model.

In this paper, we examine for the first time how predictions relevant for PTA observations are
modified by the presence of ambient gas, affecting the inspiral rate of binaries during a merger episode,
ingredient (ii) above. A gaseous envelope is expected to surround the binary because
MBHBs are produced in galaxy mergers, which are known to trigger
inflows of large quantities of gas into the central region, as shown by hydrodynamic simulations
\citep{springel}. This gas, accreted onto the MBHs, is responsible for luminous AGN activity, and
is also expected to catalyze the coalescence of the new-formed MBH pair \citep[e.g.,][]{escala04, dotti07},
as described below.
The forming MBHB spirals inward initially as a result of dynamical friction on
dark matter, ambient stars, and gas \citep{BBR}. As the binary separation shrinks to sub-pc scales,
the supply rate of stars crossing the orbit decreases, and the interaction with
stars becomes less and less efficient to shrink the binary.
In gas rich mergers, the dense nuclear gas is expected to
cool rapidly and settle into a geometrically-thin
circumbinary accretion disk \citep[e.g.,][]{barnes,escala04}.
Torques from the tidal field of the binary clear a gap in the gas with a radius
less than twice the separation of the binary, and generates a spiral density wave
in the gaseous disk, which in turn drains angular momentum away from
the binary on a relatively short timescale within the last parsec, $\lsim 10^7\,$yr
(\citealt{escala05,an02,an05,dotti07,Hayasaki09}, see however \citealt{lodato09}).
Ultimately, at even smaller separations, corresponding to an orbital timescale
of $\sim$ years, the emission of GWs becomes the dominant mechanism driving the
binary to the final coalescence.
The main point of this paper, is to notice that the most sensitive PTA frequency band
corresponds to orbital separations near the transition between gas and
GW dominated evolution . {\it As binaries shrink more
quickly inwards in the gas-driven phase, the number of binaries emitting at each given separation
is decreased compared to the purely GW-driven case.} The subject of this work
is to explore how various gas-driven models modify the expectations on the GW signal
potentially observable by PTAs.

Recently, \citet[HKM09 hereafter]{HKM09}, examined the evolution of MBHBs in the gas--driven regime
for simple models of geometrically thin circumbinary disks \citep[see also,][]{syerclarke,an05}.
The interaction between the binary and the gaseous disk is analogous to type-II
planetary migration, and evolves through two main phases. First, the inspiral
is analogous to the disk--dominated type-II migration of planetary dynamics, where
the binary migrates inwards with a radial velocity equal to that of the gas accreting
towards the center. Later, as the mass of the gas within a few binary separations becomes
less than the reduced mass of the binary, the evolution slows down, and it is analogous
to the planet--dominated (or secondary--dominated) type-II migration. In both cases, the
radial inspiral rate is still much faster than in the purely GW driven
case at orbital separations beyond a few hundred Schwarzschild radii.
For standard Shakura-Sunyaev $\alpha$--disk models,
the viscosity is assumed to be proportional to the total
pressure, and is consequently very large in the radiation pressure dominated phase at small radii,
increasing the migration rate in the radiation pressure dominated phase.
On the other hand, for $\beta$--disk models, where the viscosity is proportional to the gas pressure only, the
increase of radiation pressure does not impact the viscous timescale, and the migration
rate is relatively slower in this regime.
Finally, we note that the binary-gas interaction is also significantly different for
non--steady models of accretion (\citealt{ivanov}, HKM09). In the typical secondary--dominated phase,
gas flows in more quickly then how the binary separation shrinks, and is repelled close to the
outer edge of the gap by the torques of the binary. This causes gas to accumulate near the gap,
and delays the merger of the binary relative to the steady-state models.

In this paper, we couple the HKM09 models for the migration of MBHBs in the presence of a
steady gaseous disk, to the population models derived in SVV09, and we compute the effects on GWs
at nHz frequencies. Here, we restrict to nearly circular inspirals for simplicity, using the
corresponding GW spectrum (ingredient--iii above). This assumption might be violated in
gas driven inspirals \citep{an05,cua09}, and is the subject of a future paper
(Sesana \& Kocsis 2010, in prep).

The paper is organized as follows. In Section 2 we introduce the theory of the GW signal from a MBHB population,
describing its characterization in terms of its {\it stochastic level} and of the statistics of {\it individually
resolvable sources}. In Section 3 we describe our MBHB population model, coupling models
of coalescing binaries derived by cosmological N-body simulations to a scheme for the dynamical evolution of the
binaries in massive circumbinary disks. We present in detail our results in Section 4, and in Section 5 we briefly
summarize our main findings. Throughout
the paper we use geometric units with $G=c=1$.

\section{Description of the gravitational wave signal}
The theory of the GW signal produced by the superposition of radiation from a large number of individual
sources, was extensively presented in SVC08 and SVV09, here we review the basic concepts, deriving
the GW signal for gas driven mergers.

The dimensionless characteristic amplitude of the GW background produced by a
population of binaries with a range of masses $m_1$ and $m_2$ and redshifts $z$ is given by \citep{Phinney01}
\begin{equation}
h_c^2(f) =\frac{4}{\pi f^2}
\int \D z\D m_1\D m_2
 \, \frac{\partial^3 n}{\partial z \partial m_1 \partial m_2}
{1\over{1+z}}~{{\D E_{\rm gw}} \over {\D \ln{f_r}}}\,.
\label{e:Phinney}
\end{equation}
where $dE_{\rm gw}/d\ln{f_r}$ is the total emitted GW energy per logarithmic
frequency interval in the comoving binary rest-frame,
$f_r = (1 + z) f$ is the rest-frame frequency,
$f$ is the observed frequency, $n$ is the comoving number density
of sources, and the $1/(1+z)$ factor accounts for the redshift of the
observed GW energy. The characteristic GW amplitude $h_c$ is related to the
present day total energy in GWs as $\rho_{\rm GW}=\frac{\pi}{4}\int f h_c^2(f) \D f$.

\subsection{GW-driven inspirals}
To the leading quadrupole order, for circular purely GW-driven binaries orbiting far outside the innermost
stable circular orbit (ISCO), Eq.~(\ref{e:Phinney})
can be evaluated assuming that $E_{\rm gw}= E_{\rm pot} = -m_1m_2/a$ is equal to the Newtonian
potential energy of the binary, and that $f_r$ is equal to twice the Keplerian orbital frequency
\citep{Phinney01}. In this case
\begin{equation}
\frac{\D E_{\rm gw}}{\D \ln f_r} = \frac{1}{3} \mu (\pi M f_r)^{2/3} = \frac{1}{3}(\pi f_r)^{2/3} \Mc^{5/3}
\label{e:Egw-Phinney}
\end{equation}
for $f_r<f_{\rm ISCO}\equiv 1/(6^{3/2}\pi M)$.
Here $\mu=m_1m_2/M$, $M=m_1+m_2$, and $\Mc^{5/3}=\mu M^{2/3}$ are the reduced, total, and
chirp masses for a binary, and $f_{\rm ISCO}$ is the GW frequency at ISCO. Substituting into Eq.~(\ref{e:Phinney}),
\begin{equation}
h_c^2(f) =\frac{4 f^{-4/3}}{3\pi^{1/3}}
\int \D z\D m_1\D m_2
 \, \frac{\partial^3 n}{\partial z \partial m_1 \partial m_2}
\frac{\mu M^{2/3}}{(1+z)^{1/3}}.
\label{e:hc-GWdriven}
\end{equation}
It is also useful to examine the number of MBHBs and their respective contributions
to the total signal (\citealt{Phinney01}; SVC08). Eq.~(\ref{e:hc-GWdriven}) can be rewritten as
\begin{equation}
h_c^2(f) =
\int \D z\D m_1\D m_2
 \, \frac{\partial^4 N}{\partial m_1 \partial m_2\partial z  \partial \ln f_r}
h^2(\Mc, z, f_r),\label{e:hc_GWdrivenN}
\end{equation}
where $N$ is the number of sources, which we can calculate from a comoving merger rate density as explained below, and
\begin{equation}
h(\Mc, z, f_r) = \frac{8}{10^{1/2}} \frac{\Mc}{d_L(z)} (\pi \Mc f_r)^{2/3},
\label{e:h(f_r)}
\end{equation}
is the sky and polarization averaged characteristic GW strain amplitude of a single binary with chirp mass
$\Mc$, at the particular orbital radius corresponding to $f_r$.

We generate the distribution $\partial^4 N/(\partial m_1 \partial m_2\partial z \partial \ln f_r)$ corresponding to a
comoving merger rate density\footnote{In practice $\partial^4 N/(\partial m_1 \partial m_2 \partial t_r \partial V_c)$
is a function of $m_1$, $m_2$, and $z$.} $\partial^4 N/(\partial m_1 \partial m_2 \partial t_r \partial V_c)$ (see
Sec.~\ref{s:MergerRate}), assuming that the number of binaries emitting in the interval $\ln f_r$ is proportional
to the time the binary spends at that frequency,
\begin{equation}
\frac{\partial^4 N}{\partial m_1 \partial m_2\partial z  \partial \ln f_r} =
\frac{\partial^4 N}{\partial m_1 \partial m_2 \partial t_r \partial V_c} \frac{\D V_c}{\D z}
\frac{\D z}{\D t_r} \frac{\D t_r}{\D \ln f_r}
\end{equation}
where  $\D V_c/\D z$ and $\D z/\D t_r$ are given by the standard cosmological relations between comoving volume,
redshift, and time, given in e.g. \cite{Phinney01}. The last factor can be expressed using the  residence time
$\tres = a (\D a/\D t_r)^{-1}$ the binary spends at a particular semimajor axis as
\begin{equation}
\left|\frac{\D t_r}{\D \ln f_r}\right|
= \left|\frac{\D t_r}{\D \ln a}\frac{\D \ln a}{\D \ln f_r}\right|
= \frac{2}{3}\tres,
\label{e:dtdlnf}
\end{equation}
where  Kepler's law, $a = M (\pi M f_r)^{-2/3}$, was used to obtain $\D \ln a/\D \ln f_r = 2/3$, and the
residence time for a purely GW-driven evolution is
\begin{equation}
\tres = \tres^{\gw} \equiv \frac{\D t_r}{\D \ln a} = \frac{5 }{64} \Mc (\pi \Mc f_r)^{-8/3}\label{e:tresgw}.
\end{equation}
In summary, the distribution of sources in Eq.~(\ref{e:hc_GWdrivenN}) becomes
\begin{equation}
\frac{\partial^4 N}{\partial m_1 \partial m_2 \partial z \partial \ln f_r}
=
\frac{2}{3}\frac{\partial^4 N}{\partial m_1 \partial m_2 \partial t_r \partial V_c} \frac{\D V_c}{\D z}
\frac{\D z}{\D t_r} \tres.\label{e:hc_GWdrivenN2}
\end{equation}
Equations (\ref{e:hc-GWdriven}) and (\ref{e:hc_GWdrivenN}--\ref{e:hc_GWdrivenN2}) are equivalent, but
(\ref{e:hc_GWdrivenN}--\ref{e:hc_GWdrivenN2}) are practical to generate discrete Monte Carlo realizations of
a given source population. Moreover, Equations~(\ref{e:hc_GWdrivenN}--\ref{e:hc_GWdrivenN2}) provide a
transparent interpretation of (\ref{e:hc-GWdriven}). The total RMS background, $h_c \propto \sqrt{N} h \propto f^{-2/3}$,
comes about because the mean number of binaries per frequency bin is $N\propto \tres^{\gw} \propto f_r^{-8/3}$
and each binary generates an RMS strain $h\propto f_r^{2/3}$. The scaling $h_c\propto f^{-2/3}$ is a consequence
of averaging over the local inspiral episodes of merging binaries in the GW driven regime, but is completely
independent on the overall cosmological merger history or on the involved MBH masses. The latter affects only
the overall constant of proportionality. Eq.~(\ref{e:hc-GWdriven}) also shows that this scaling constant is
insensitive to the cosmological redshift distribution of mergers,  $h_c\propto(1+z)^{1/6}$, as well as the
number of minor mergers with $\mu\ll M$, once the total mass in satellites merging with BHs of mass $M$ is
fixed \citep{Phinney01}. However, the background is sensitive to the assembly scenarios of major mergers (SVV09).
More importantly, as shown in SVC08, the actual GW signal in any single realization of inspiralling binaries is
qualitatively different from $h_c\propto f^{-2/3}$, as a small discrete number of individual
massive binaries dominate the nHz GW-background, creating a very spiky GW spectrum. We further discuss the
discrete nature of the signal in Section~\ref{s:statistics} below.

\subsection{Gas-driven inspirals}
\label{s:gas}
Let us now derive the GW background for an arbitrary model of binary evolution. We can derive the background
using the residence time $\tres$ the binary spends at each semimajor axis $a$, where $\tres < \tres^{\gw}$ in
the gas driven phase (cf. Eq.~[\ref{e:tresgw}] for the purely GW driven case). We define $\tres$ for various
accretion-disk models in Section~\ref{s:gasdetails}. Generally, the emitted GW spectrum is
\begin{equation}
\frac{\D E_{\rm gw}}{\D \ln f_r} = \frac{\D E_{\rm gw}}{\D t_r}  \frac{\D t_r}{\D \ln f_r} =
\frac{2}{3} \frac{\D E_{\rm gw}}{\D t_r}\tres.
\end{equation}
The second equality follows the definition of $\tres$ and Kepler's law (see Eq.~[\ref{e:dtdlnf}]). The emitted
power $\D E_{\rm gw}/\D t_r$ depends only on the masses and the geometry of the orbit, but is independent of the
global migration rate of the binary, and therefore it is the same as in the pure GW driven case. The effects of
migration is fully encoded in $\tres$. Plugging back in Eq.~(\ref{e:Phinney}), the mean square signal in the gas
driven phase is
\begin{eqnarray}
h_c^2(f) =&&\frac{4 f^{-4/3}}{3\pi^{1/3}}
\int \D z\D m_1\D m_2
 \, \frac{\partial^3 n}{\partial z \partial m_1 \partial m_2}
\frac{\mu M^{2/3}}{(1+z)^{1/3}} \nonumber\\
&&\times \frac{\tres(M, \mu, f_r)}{\tres^{\gw}(\Mc, f_r)}
\label{e:hc-gasdriven}
\end{eqnarray}
A comparison with (\ref{e:hc-GWdriven}) shows that the RMS signal is attenuated in the gas driven phase by
$\sqrt{\tres(M, \mu, f_r)/\tres^{\gw}(\Mc, f_r)}$. This factor is a complicated function of $f_r$, that behaves
differently for different masses and mass ratios of the binaries; the overall spectrum is no longer a powerlaw.

We generate Monte Carlo realizations of the GW signal by sampling the population of the inspiralling systems.
To do this, it is sufficient to recognize that the derivation given by Eqs.~(\ref{e:hc_GWdrivenN}--\ref{e:hc_GWdrivenN2})
remains valid in the gas drive phase if using the appropriate $\tres(M, \mu, f_r)$, since the contribution given
by each individual source to the signal is the same. {\it The net GW spectrum changes because of a reduction in
the number of sources in the gas driven case}.

Note that the individual GW signal given by Eq.~(\ref{e:h(f_r)}) depends on three parameters only:
$\Mc$, $z$, and $f_r$. This implies that signals from sources with the same observed frequency, redshift, and chirp
mass, $(f_r, z, \Mc)$, but different mass ratios, $q$, are totally indistinguishable\footnote{This is true only
in the angular averaged approximation. We neglect the directional sensitivity of PTAs.}. We can make use of this
property and reduce the number of independent parameters in the distribution by integrating over the mass ratio
in Eq.~(\ref{e:hc_GWdrivenN})
\begin{equation}
\frac{\partial^3 N}{\partial \Mc \partial z \partial \ln f_r}
=
\int_0^1 dq
\frac{\partial^4 N}{\partial m_1 \partial m_2 \partial z \partial \ln f_r}\left|\frac{\partial(m_1,m_2)}{\partial(\Mc,q)}\right|,
\label{e:hc_GWdrivenN3}
\end{equation}
Note, that this step is different for the GW and gas driven cases, because
$\partial^4 N/(\partial m_1 \partial m_2 \partial z \partial \ln f_r)$ is proportional to $\tres$ in
Eq.~(\ref{e:hc_GWdrivenN2}). Here $|\partial(m_1,m_2)/\partial(\Mc,q)|$ is the determinant of the
Jacobian matrix corresponding to the variable change from $(m_1, m_2)$ to $(\Mc, q)$. With Eq.~(\ref{e:hc_GWdrivenN2}) and
(\ref{e:hc_GWdrivenN3}) we derive $\partial^3 N/(\partial \Mc \partial z \partial \ln f_r)$ for
any gas driven model given by $\tres(M,\mu,f_r)$, and draw Monte Carlo
samples of inspiralling binaries from this distribution when generating the GW signal.

\subsection{Statistical characterization of the signal}
\label{s:statistics}

In observations with PTAs, radio-pulsars are monitored weekly for total periods of
several years.
Assuming a repeated observation in uniform $\Delta t$ time intervals for a total time $T$,
the maximum and minimum resolvable frequencies are $f_{\max}=1/(2\Delta t)$, corresponding to
the Nyquist frequency, and $f_{\min}= 1/T$. The observed GW spectrum is therefore discretely sampled
in bins of $\Delta f = f_{\min}$. For circular orbits, the frequency of the GWs is twice the orbital
frequency.

Let us examine whether the sources' GW frequency evolves during the observation relative to the size of the
frequency bins.  Writing the frequency shift during an observation
time $T$ as $\Delta f_{\rm evol} \approx \dot{f}T = (\D \ln f/\D \ln a) (\D \ln a/ \D t) f T = \frac{3}{2}  f T/ [(1+z)\tres]$ and considering the frequency resolution bin to be $\Delta f_{\rm bin}=1/T$,
then the frequency evolution relative to the frequency resolution bin is
\begin{equation}\label{e:fevolution}
 \frac{\Delta f_{\rm evol}}{\Delta f_{\rm bin}}\approx \frac{3}{2} \frac{ f T^2}{(1+z) \tres}
 =  \frac{0.015}{1+z}{\Mc}_{8.5}^{5/3} f_{50}^{11/3}T_{10}^2 \frac{\tres^{\gw}}{\tres},
\end{equation}
where ${\Mc}_{8.5}$ is the chirp mass in units of $10^{8.5}\Ms$, $f_{50}$ is the frequency in units of 50 nHz,
$T_{10}$ is the observation time in unit of 10 years, and for the second equality we have used equation~(\ref{e:tresgw}).
Equation~(\ref{e:fevolution}) shows that
typical binaries contribute to a single frequency bin as stationary sources in the GW-driven regime.\footnote{The detection of
the frequency shift of an ${\Mc}_{8.5}=f_{50}=z=1$ source  would require an extended observation with $T\gsim 35$\,yr.}
We shall demonstrate that this is also true in the gas driven case (see Figure~\ref{figtres} below).

We generate $N_k$ different realizations of the signal (usually $N_k=1000$), i.e. $N_k$ realizations of the MBHB
population, consistent with Eq.~(\ref{e:hc_GWdrivenN2}) (see Section~\ref{s:MergerRate}). Each of those consists
of $N_b\sim 10^3-10^4$ binaries producing a relevant contribution to the signal, which we label by
$({\Mc}_i, z_i, {f_r}_i)$,  $i=1,2,\dots,N_{\rm b}$. The total signal (Eq.~[\ref{e:hc_GWdrivenN}]) in each frequency
resolution bin $\Delta f$ is evaluated as the sum of the contributions of each individual source
(see \cite{pau09} for the detailed numerical procedure)
\begin{equation}
h_c^2(f) =
\sum_i h_{c,i}^2({\Mc}_i, z_i, {f_r}_i),\label{e:hc_Nsum},
\end{equation}
where for each $f$ the sum is over the inspiralling sources emitting in the corresponding blue--shifted (i.e. restframe)
$\Delta f_r$ frequency resolution bin, and $h_{c,i}=h_i\sqrt{f_iT}$ is the angle and polarization averaged GW strain given
by Eq.~(\ref{e:h(f_r)}), multiplied by the square root of the number of cycles completed in the observation time. For comparison, we also
evaluate the continuous integral Eq.~(\ref{e:hc_GWdrivenN}), which represents the RMS average of Eq.~(\ref{e:hc_Nsum})
over different realizations of MBHB populations in the limit $N_k\rightarrow\infty$.

Since the mass function of merging binaries is in general quite steep,
the relative contributions of the few most massive binaries turn out to dominate the
background in each frequency bin. The total GW signal depends very sensitively on
these rare binaries, and the inferred spectrum is very spiky.
It is useful to separate the total signal $h_c$ into
a part generated by a population of GW--bright {\it individually resolvable sources},
and a {\it stochastic level} $h_{s}$, which
includes the contribution of all the unresolvable, dimmer sources. More precisely, in each frequency resolution bin,
we find the MBHB with the largest $h^2_{c,i}({\Mc}_i, z_i, {f_r}_i)$, and we define it {\it individually resolvable}
if its signal is stronger than the total contribution of all the other sources in that particular frequency bin. The
{\it stochastic level} is consequently defined by adding up only the unresolvable sources in Eq.~(\ref{e:hc_Nsum}).
Since the signal is dominated by few individual sources in each frequency bin, the $h_s(f)$ distribution obtained over
the $N_k$ realizations is far from being Gaussian or just even symmetric. To give an idea of the uncertainty range of
$h_s(f)$, we calculate the $10\%$, $50\%$ and the $90\%$ percentile levels of the $h_s(f)$ distribution of the $N_k$
different realizations.

The separation of individually resolvable sources is useful for several reasons (see SVC; SVV09).
First, it is useful from a statistical point of view for understanding the variance of
the expected GW spectrum among various realizations of the inspiralling MBHBs.
The discrete nature of the resolvable sources allows a different statistical analysis than for
the smooth background level corresponding to the stochastic level, $h_s(f)$.
The individually resolvable signals could also be important observationally.
A sufficiently GW--bright resolvable binary allows to measure the GW polarization using PTAs,
and give information on the sky position of the binary \citep{SV10},
which might be used to search for direct electromagnetic signatures
like periodically variable AGN activity (HKM09).
A coincident detection of GWs and electromagnetic emission of the same
binary  would have far reaching consequences in fundamental physics, cosmology, and black hole physics
\citep{Kocsis08}.

\subsection{Timing Residuals}
\label{s:timing}
In general, the characteristic GW amplitudes (either of a stochastic background or of a resolvable source) can be
translated into the pulsar timing language by converting $h_c(f)$ into a ``characteristic timing residual'' $\delta t_c(f)$
corresponding to the sky position and polarization averaged delay in the time of arrivals of consequent pulses due to GWs,
\be
\delta t_c(f)=\frac{h_c(f)}{2\pi f}.
\label{e:t_c}
\ee

The pulsar timing residuals expected from an individual stationary GW source is derived in section 3
of SVV09 in detail. The corresponding measurement can be represented, in the time domain, with a residual:
\be
\delta t(t) = r(t) + \delta t_{\rm N}(t)
\ee
where $r(t)$ is the contribution due to the GW source (which accumulates continuously with observing time $t$, see below),
and $\delta t_{\rm N}(t)$ represents random fluctuations due to noise.
The latter is the superposition of the intrinsic noise in the measurements and the GW stochastic level
from the whole population of MBHBs, with a root-mean-square (RMS) value
\be
\delta t_\mathrm{N,rms}^2(f) = \langle \delta t_\mathrm{N}^2(f) \rangle = \delta t_\mathrm{p}^2(f) + \delta t_\mathrm{s}^2(f).
\label{e:noise}
\ee
where $t_\mathrm{p}(f)$ is the RMS instrumental and astrophysical noise corresponding to
the given pulsar, and $\delta t_\mathrm{s}(f)=h_s(f)/(2\pi f)$ is due to the RMS stochastic GW background of unresolved MBHBs,
as defined in the previous section.

The sky angle and orbital orientation-averaged signal-to-noise ratio (SNR) at which one MBHB, radiating at (GW)
frequency $f$, can be detected using a {\it single} pulsar with matched filtering is
\be
{\rm SNR}^2 = \frac{\delta t_\mathrm{gw}^2(f)}{\delta t_\mathrm{N,rms}^2(f)}\,.
\label{e:snr}
\ee
Here $\delta t_\mathrm{gw}(f)$ is the root-mean-squared timing residual signal resulting from GWs emitted by the
individual stationary source over the observation time $T$ defined as:
\be
\delta t_\mathrm{gw}(f) =\sqrt{\frac{8}{15}}\frac{h(\Mc, z, f)}{2\pi f} \sqrt{f T}
\label{e:deltatgw}
\ee
where $h(\Mc, z, f)$ is the angle and polarization averaged GW strain amplitude given by Eq.~(\ref{e:h(f_r)}), the
prefactor $\sqrt{8/15}$ averages the observed signal over the ``antenna beam pattern'' of the array (Eq. (21)
in SVV09\footnote{Note that that the square root in the prefactor $\sqrt{8/15}$ is missing in Eq. (20) of SVV09
because of a typo there.}), and the $\sqrt{fT}$ term accounts for the residual build-up with the number of cycles.
For $N_{\rm p}$ number of pulsars, the total detection SNR, of an individually resolvable MBHB is the RMS of the
contributions of individual pulsars given by (\ref{e:snr}). For $N_{\rm p}$ identical pulsars, the effective
noise level is therefore attenuated by $N_{\rm p}^{-1/2}$.

In the following we will represent the overall GW signal and the stochastic background by using either their
characteristic amplitudes, $h_c(f)$ and $h_s(f)$, or the corresponding characteristic timing residuals,
$\delta t_c(f)$ and $\delta t_s(f)$, according to equation (\ref{e:t_c}). We study the detection significance of individually
resolvable sources and the distribution of their numbers as a function of the induced $\delta t_{\rm gw}$. For each Monte Carlo
realization of the emitting MBHB population, we count the cumulative number of all ($N_t$) and resolvable ($N_r$) sources
above $\delta t_{\gw}$ as a function of $\delta t_{\rm gw}$:
\be
N_{t/r}(\delta t_\mathrm{gw}) = \int_{\delta t_\mathrm{gw}}^\infty \frac{\partial N_{t/r}}{\partial \delta t_\mathrm{gw}'}\delta t_\mathrm{gw}'\,,
\label{e:N}
\ee
where the integral is either over all sources or only the individually resolvable sources (i.e. restricted to those
that produce residuals above the RMS stochastic level, see Sec.~\ref{s:statistics}).

\section{The emitting binary population}
\label{s:Population}
We calculate the GW signal by generating a catalogue of binaries consistent with
Eq.~(\ref{e:hc_GWdrivenN2}). This requires (i) a model for the
comoving merger rate density of coalescing MBHBs,
$\partial^4 N/(\partial m_1 \partial m_2 \partial t_r \partial V_c)$, and
(ii) a model for the evolution of individual inspiralling binaries $\tres(M, \mu, f_r)$.
These two items are the subjects of the next two subsections below.

\subsection{Population of coalescing massive black hole binaries}
\label{s:MergerRate}
We use the population models described in section 2 of SVV09,
the reader is deferred to that paper for full
details. We extract catalogs of merging binaries from the semi-analytical
model of \cite{Bertone07} applied to the Millennium run \citep{springel}.
We then associate a central
MBH  to each merging galaxy in our catalogue.
We explored a total of 12 models, combining four $M_{\rm BH}$-bulge
prescriptions found in the literature with three different accretion scenarios
during mergers. The twelve models are listed in table 1 of SVV09.
In the present study we shall use the Tu-SA population model as our default case.
In this model, the MBH masses in the merging galaxies correlate with the
masses of the bulges following the relation reported in \cite{Tundo07},
and accretion is efficient onto the more massive black hole, {\it before}
the final coalescence of the binary.
Since the comparison among different MBHB population models is not the main purpose
of this study, we will present results only for this model. However, we tested our
dynamic scenario on other population models presented in SVV09,
finding no major differences for alternative models.

Assigning a MBH to each galaxy, we obtain a catalogue of mergers
labelled by MBH masses and redshift. From this, we generate the merger rate per comoving volume,
$\partial^4 N/(\partial m_1 \partial m_2 \partial t_r \partial V_c)$. In practice, due to the
large number of mergers in the simulation, this is a finely resolved continuous function,
describing the merger rate density as a function of $m_1$, $m_2$, and $z$.
After plugging into Eqs.~(\ref{e:hc_GWdrivenN2})
and (\ref{e:hc_GWdrivenN3}), we can obtain the continuous distribution
$\partial^3 N/(\partial \Mc \partial z \partial \ln f_r)$ one would observe in an ``ideal snapshot'' of the
whole sky. We then sample this distribution to generate random Monte Carlo realizations
of the GW signal. In summary, the
chosen MBHB population model fixes the cosmological merger history, i.e. the function
$\partial^4 N/(\partial m_1 \partial m_2 \partial t_r \partial V_c)$, and ``Monte Carlo
sampling'' refers to first choosing a realization of the cosmological merger history
(i.e. generate $N_b$ number of binary masses and redshifts\footnote{Here $N_b$ is chosen randomly
for each distribution, it can vary for different realizations of the population according to a Gaussian
distribution with $\sigma=1/\sqrt{\langle N_b\rangle}$ around the mean $\langle N_b\rangle$.})
 and then assigning an orbital frequency (or time to merger) to each binary.
In addition to our fiducial merger rate, we also examine for the first time the situation where minor mergers
do not contribute to the coalescence rate of the central MBHs. This is motivated by recent
numerical simulations indicating that minor mergers with mass ratios $q<0.1$ lead to tidal
stripping of the merging satellite, and the resulting core does not sink efficiently to the
centre of the host galaxy \citep{cal09}. We identify the mass ratio of merging galaxies using the mass
of the stellar components, to avoid complications due to the tidal stripping of merging dark matter
halos.
In practice, we find that suppressing all the minor mergers does not affect the resulting GW signal,
implying that the contribution of mergers involving dwarf galaxies is negligible.

\subsection{Binary evolution in massive circumbinary disks}
\label{s:gasdetails}
We adopt the simple analytical models of HKM09 to describe the dynamical evolution
of MBHBs in a geometrically thin circumbinary accretion disk. {Here we provide 
only a short summary of the gas-driven models, highlight their main assumptions, 
and defer the reader to section 2 of HKM09 and references therein for more details. 

For the typical MBH masses 
and separations we are considering, the tidal torque from the binary dominates
over the viscous torque in the disk, opening a gap in the gas distribution. A spiral density
wave is excited in the disk which torques the binary and pushes it inward. First, when the
binary separation is relatively large, and the local disk mass is larger than the mass
of the secondary, then the 
secondary migrates inward with the radial velocity of the accreting gas, 
\begin{equation}\label{e:tnu}
 \tres^{\nu} = -\frac{a}{\dot{a}_{\nu}} = \frac{2\pi r_0^2 \Sigma_0}{\dot M}
\end{equation}
where $r_0$ is the radius of the gap (typically $r_0\sim 1.5 a$ where $a$ is the semimajor
axis of the binary), $\Sigma_0$ is the local surface density of the disk with no secondary, 
and $\dot M$ is
the accretion rate far from the binary (the $\nu$ index denotes viscous evolution).
This is analogous to disk dominated Type-II planetary migration.
More typically, however, the binary is more massive than the local 
disk mass. In this case, analogous to secondary-dominated Type-II migration, the
angular momentum of the binary is absorbed less efficiently by the gas outside the gap, 
and the evolution slows down according to \citep{syerclarke}
\begin{equation}\label{e:tSC}
 \tres^{\rm SC} = -\frac{a}{\dot{a}_{\rm SC}} = q_B^{-k_1} t_{\nu}
\end{equation}
where $q_B$ is a measure of the lack of local disk mass dominance
\begin{equation}
 q_B = \frac{4\pi r_0^2 \Sigma_0}{\mu} = \frac{2 \dot M}{\mu} t_\nu,
\end{equation}
which is less than unity in this case; 
$\mu$ is the reduced mass of the binary, 
and $k_1$ is a constant that depends on the surface density -- accretion rate 
powerlaw index. Here $k_1=0$ if $q_B>1$; otherwise, 
$k_1=7/17$ at large separations, if the disk opacity at the gap boundary is dominated by free-free absorption,
and $k_1=3/8$ closer in, if the opacity is dominated by electron scattering. 

{ In the secondary-dominated regime, eq.~(\ref{e:tSC}) assumes that
the migration is very slow, so that the binary can influence the surface density very far 
upstream, and an approximate steady state is reached where the gas density 
is enhanced outside the gap by a factor $q_B^{-k}$ 
relative to the gas density with no secondary \citep{syerclarke}. To examine the
sensitivity of our conclusions to these assumptions, we consider an additional 
nonsteady model of \citet{ivanov}. In their model, which is also axisymmetric by
construction, they assumed that the binary torques are 
concentrated in a narrow ring near the edge of the cavity, which results in a
time-dependent pile up of material near the cavity. 
The migration slows down further relative to $\tres^{\rm SC}$ as
\begin{equation}\label{e:tIPP}
 \tres^{\rm IPP} = \frac{\mu}{2 \dot M} \left(\frac{a}{a_0}\right)^{1/2} 
\left\{\frac{1}{1 + \delta \left(1 - \sqrt{a/a_0}\right)}\right\}^{k_2},
\end{equation}
where $\dot M$ is the constant accretion rate far outside the binary, 
$a_0$ is the initial semimajor axis where the local disk mass is just
equal to the secondary (i.e. $q_B=1$), $\delta$ and $k_2$ are
constants which depend on the density--accretion rate profile of the disk 
($\delta =4.1$ and $6.1$, $k_2=0.29$ and  $0.26$ for free-free and electron scattering opacity, 
respectively). { Note that as the separation decreases well below its initial value, $a\ll a_0$, 
the curly bracket is $a$--independent,  
implying that $\tres^{\rm IPP} \propto a^{1/2} \propto \torb^{1/3}$.}

Since the migration rate is proportional to the gas surface density and accretion rate, it is 
sensitive to the structure of the accretion disk. Following \citet{syerclarke},
we estimate $\Sigma_0$ with that of a steady accretion disk of a single accreting 
BH, but consider two different models for that, $\alpha$ and $\beta$--disks (see HKM09 for 
explicit formulae). }
For the classic \cite{ss73} $\alpha$-disk,
the viscosity is proportional to the total (gas+radiation) pressure of the
disk.
Until very recently, this model, if radiation pressure dominated,
has been thought to be thermally and viscously unstable \citep{le74,piran78}.
In the alternative $\beta$-model, the viscosity
is proportional to the gas pressure only\footnote{The name comes from the
definition of viscosity $\nu\propto \alpha p_{\rm gas}^{\beta}p_{\rm tot}^{1-\beta}$
where $\beta=1$ for the $\beta$--model, while $\beta=0$ for the
$\alpha$ model. In both cases $\alpha$ is a free model parameter.}, and
it is stable in both sense. The nature of viscosity is not well understood to predict
which of these prescriptions lies closer to reality.
Recent numerical magneto-hydrodynamic simulations \citep{hirose09}
suggests that the thermal instability is avoided in radiation pressure
dominated situations because stress fluctuations lead
the associated pressure fluctuations, and seem to favor the $\alpha$
prescription over the $\beta$--model.
We carry out all calculations for both models, but consider the $\alpha$ prescription
as our fiducial disk model.
In both cases, the model is uniquely determined by three parameters:
the central BH mass, the accretion rate { far from the binary} $\dot M$, and
the $\alpha$ viscosity parameter. The exact value of these parameters
is not well known. Observations of luminous AGN
imply an accretion rate around $\dot m = \dot{M}/\dot{M}_{\rm Edd} = (0.1$--$1)$
 with a statistical increase towards higher quasar luminosities \citep{Kollmeier06,Trump09}.
Here $\dot{M}_{\rm Edd}=L_{\rm Edd}/(\eta{\rm c}^2)$
is the Eddington accretion rate for $\eta=10\%$ radiative efficiency, where
$L_{\rm Edd}$ is the Eddington luminosity. Observations of outbursts in binaries
with an accreting white dwarf, neutron star, or stellar black hole imply
$\alpha=0.2$--$0.4$ \citep[and references therein]{Dubus01,King07}.
Theoretical limits based on simulations of magneto-hydrodynamic turbulence
around black holes are inconclusive, but are consistent with
$\alpha$ in the range 0.01--1 \citep{Pessah07}. It is however unclear whether
these numbers are directly applicable to circumbinary MBH systems.
The binary exerts a torque that pushes the gas away on average, and 
consequently  might reduce the accretion rate (see \citealt{Lubow99,LubowDAngelo06}, 
in the planetary context, and \citealt{mm08,cua09}, for MBH binaries).
We explore several choices covering all 6 combinations with
$\dot{m}=\{0.1, 0.3\}$ and $\alpha=\{0.01, 0.1, 0.3\}$ for both $\alpha$ and $\beta$--disks,
respectively. Motivated by the considerations outlined above,
we highlight the $\alpha$-disk with $\dot{m}=0.3$ and $\alpha=0.3$ as our default
model.

{ We regard $\tres^{\rm SC}$ and $\tres^{\rm IPP}$ as lower and upper limits of the
true residence time during secondary dominated gas driven migration. However, 
we caution that these estimates are subject to many uncertainties related to 
the complexity of binary accretion disk and migration physics. 
Other models for the binary--disk interaction \citep[i.e.][]{Hayasaki09,Hayasaki10}
lead to a slower migration rate.
Two dimensional hydrodynamic simulations show that the flow into the gap is 
generally non-axisymmetric, and the gap and the binary orbit becomes eccentric
\citep{LubowDAngelo06,mm08,cua09}. The accretion rate and inspiral velocity
in reality may be higher than in the axisymmetric approximation. 
Resonant interactions may lead an enhanced angular momentum transport 
\citep{gt79}. For unequal masses or thick disks, there may be 
significant inflow into the gap. In this case, corotation torques need to be considered,
and the migration is expected to be much faster inward or outward, analogous to
Type I planetary migration \citep{ttw02}. Disk thermodynamics 
may also significantly influence the migration rate \citep{Paardekooper1,Paardekooper2}. }

These steady state accretion disk models also assume that the self-gravity of the disk is 
negligible, the disk is optically thick, and the temperature of the gas is large enough
($T\gsim 10^4\,$K) that the opacity corresponds to that of ionized gas. These 
assumptions break down at radial distances corresponding to orbital periods of 
around $\torb\gsim(5$--$10)\yr$ for BHs with masses $M\gsim (10^8$--$10^9)\msun$
(see Fig. 1 of HKM09). Thus, these disk models are self-consistent for the 
relatively short period binaries, but become suspect for binaries emitting at 
the low frequency edge of the PTA window.

Another shortcoming of these gas-driven migration models, is that they were derived assuming
the disk surface density is a decreasing function of radius. This affects the specific values of $k_1$ and $k_2$ 
in eqs.~(\ref{e:tSC}) and (\ref{e:tIPP}). This assumption is safe for $\beta$--disks in general, as well as for 
$\alpha$--disks if the gas pressure dominates over radiation pressure, but it is violated for radiation
pressure dominated $\alpha$--disks.{ However,
gas-driven models with binary separation in the PTA frequency band and large masses $M>10^8\msun$ 
are in fact radiation pressure dominated near the gap.\footnote{Note that the most massive binaries, however, are GW driven, and are not sensitive to the accretion disk. } }
It is unclear how $k_1$ and $k_2$ change for radiation pressure dominated $\alpha$--disks.
As an approximation, following \citet{an05} and HKM09, we extrapolate the values of the gas pressure 
dominated outer regions.

\begin{figure}
\centering
\mbox{\includegraphics[width=84mm]{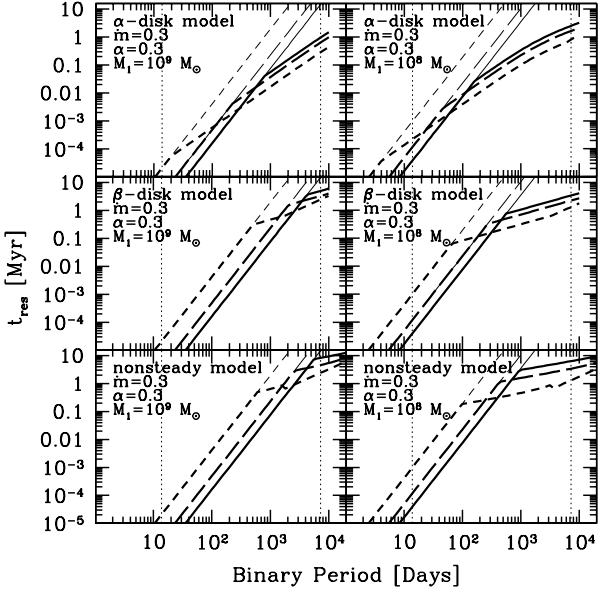}}
\caption{Residence time, $t_{\rm res}$, as a function of the binary period. Different
panels correspond to
selected model parameters as labelled.
In each panel the thick curves represent the residence time for the
gas driven dynamics according to HKM09.
Solid, long--dashed and short--dashed curves are for mass ratios $q=1, 0.1, 0.01$ respectively. The
two thin dotted vertical lines approximately bracket the PTA observable window.
The extrapolated pure GW-driven evolution ($t_{\rm res}^{\rm gw}\propto t_{\rm orb}^{8/3}$)
is shown as thin lines, for comparison.
The ratio $t_{\rm res}/t_{\rm res}^{\rm gw}$ gives the relative
decrease in the number of binaries in the gas dominated case compared to the GW driven
case.
}
\label{figtres}
\end{figure}

{
Substituting the surface density profiles $\Sigma_0(r)$ of 
$\alpha$ and $\beta$--disks in eqs.~(\ref{e:tnu}--\ref{e:tSC}), 
we obtain the inspiral rate of binaries as they evolve
from large to small radii in a gaseous medium. Figure~\ref{figtres} shows the 
corresponding residence times for selected MBH masses
and disk models. The binary migration history, can be divided in three phases.
\begin{enumerate}
 \item Initially, migration occurs on the viscous timescale $\tres^{\nu} \propto \torb^{5/6}$.
 \item Then as the disk mass decreases below the secondary mass, the gas-driven migration slows 
down according to $\tres^{\rm SC}$ or $\tres^{\rm IPP}$. The migration rate in this case is nonuniform, 
it changes as the local accretion disk structure varies due to changes in the dominant source of opacity
(free-free for large separation to electron-scattering for smaller separations) and 
the source of pressure (thermal gas pressure at large $a$ to radiation pressure at small $a$).
In particular for the steady \citet{syerclarke} models, $\tres^{\rm SC}\propto \torb^{25/51}$--$\torb^{7/12}$ 
with free-free to electron scattering opacity for thermal pressure support, and 
$\tres^{\rm SC}\propto \torb^{35/24}$--$\torb^{7/12}$, 
for radiation pressure support for $\alpha$ to $\beta$ disks, respectively; 
while for the time-dependent models of \citet{ivanov}, the accretion rate approaches $\tres^{\rm IPP} \propto \torb^{1/3}$.
Note that the residence time in this regime is generally smaller for $\alpha$--disks than for $\beta$--disks. 
This is explained by the fact that the overall surface density is smaller for radiation pressure 
dominated $\alpha$--disks, and so to achieve the same net accretion rate, the radial gas inflow
velocity is larger for $\alpha$--disks. The migration rate is a monotonic function of the radial
inflow velocity (see eqs.~\ref{e:tnu} and \ref{e:tSC}), so that the binary is pushed in more 
quickly for $\alpha$--disks.  
 \item Finally, when the separation is sufficiently reduced, the emission of gravitational
waves becomes very efficient and determines the inspiral rate according to
$\tres^{\rm gw}=a/\dot a_{\rm gw} \propto a^4  \propto \torb^{8/3}$
(see eq.~\ref{e:tresgw}). 

\end{enumerate}

}

Note that { for circular orbits, the GW frequency (in the binary restframe)} is simply $f_r = 2/\torb$.
The number of binaries at any given $\torb$ is proportional to the residence time $\tres$. Therefore,
the decrease in $\tres$ in the gas driven regime compared to the GW driven case implies a decrease
in the population of MBHs, which ultimately leads to the attenuation of
the low frequency end of the observable total GW spectrum. The RMS GW spectrum averaged
over the whole population of binary inspiral episodes is no longer a powerlaw.
It is interesting to examine the contributions of
various evolutionary phases according to Eq.~(\ref{e:hc-gasdriven}),
$h_c\propto f^{-2/3} \sqrt{\tres/\tres^{\rm gw}}$. If all binaries were
in the GW driven phase then $h_c^{\rm gw}\propto f^{-2/3}$. If all were in
the secondary-dominated Type-II migration regime with a steady radiation pressure dominated disk\footnote{
Most of the gas driven binaries contributing to the background at large frequencies are in this regime.}
$h_c^{\rm SC}\propto f^{-1/16}$--$f^{3/8}$ for $\alpha$ and $\beta$ disks, respectively,
further out $h_c^{\rm SC}\propto f^{3/8}$--$f^{43/102}$ in the gas pressure dominated regime
(with electron scattering versus free-free opacity), { while  $h_c^{\rm IPP}\propto f^{1/2}$ asymptotically for the 
nonsteady models}, and finally
$h_c^{\nu}\propto f^{1/4}$ in the disk-dominated Type-II migration regime. In any case, the
GW spectrum $h_c(f)$ is much shallower in the gas driven phase.
Note the gas driven phase contributes a nearly flat or an
{\it increasing} spectrum $h_c(f)$, very different from the
nominal $f^{-2/3}$ GW driven case. { In general, the orbital separation for more massive objects at a given $f_r$ 
is smaller in terms of Schwarzschild radii. Since the transition to gas driven migration}, when expressed in
Schwarzschild radii, is roughly independent of mass, it follows then at any given frequency bin the more massive
objects are typically GW driven and lighter objects are gas driven. The total average spectrum\footnote{if averaging
over each binary episode but not over the cosmological merger tree} assuming only wet mergers\footnote{Here 
``wet'' refers to gas rich mergers where the dynamics of the binary { is driven by both the circumbinary disk 
and GW emission depending on the binary separation, while ``dry'' refers to the case where the binary dynamics 
is driven by GW emission only at all radii.} Note that in both cases we neglect interactions with stars.},
is between $f^{-2/3}$ and $f^{1/2}$ depending on the ratio of GW driven to gas driven binaries.
Note however, that the individually resolvable
sources are typically very massive and are in the GW-driven phase for the relevant range of binary separations, and
therefore their properties should not be modified by gas effects.

\section{Results}

\subsection{Description of the signal}
   \begin{figure*}
   \centering
   \resizebox{\hsize}{!}{\includegraphics[clip=true]{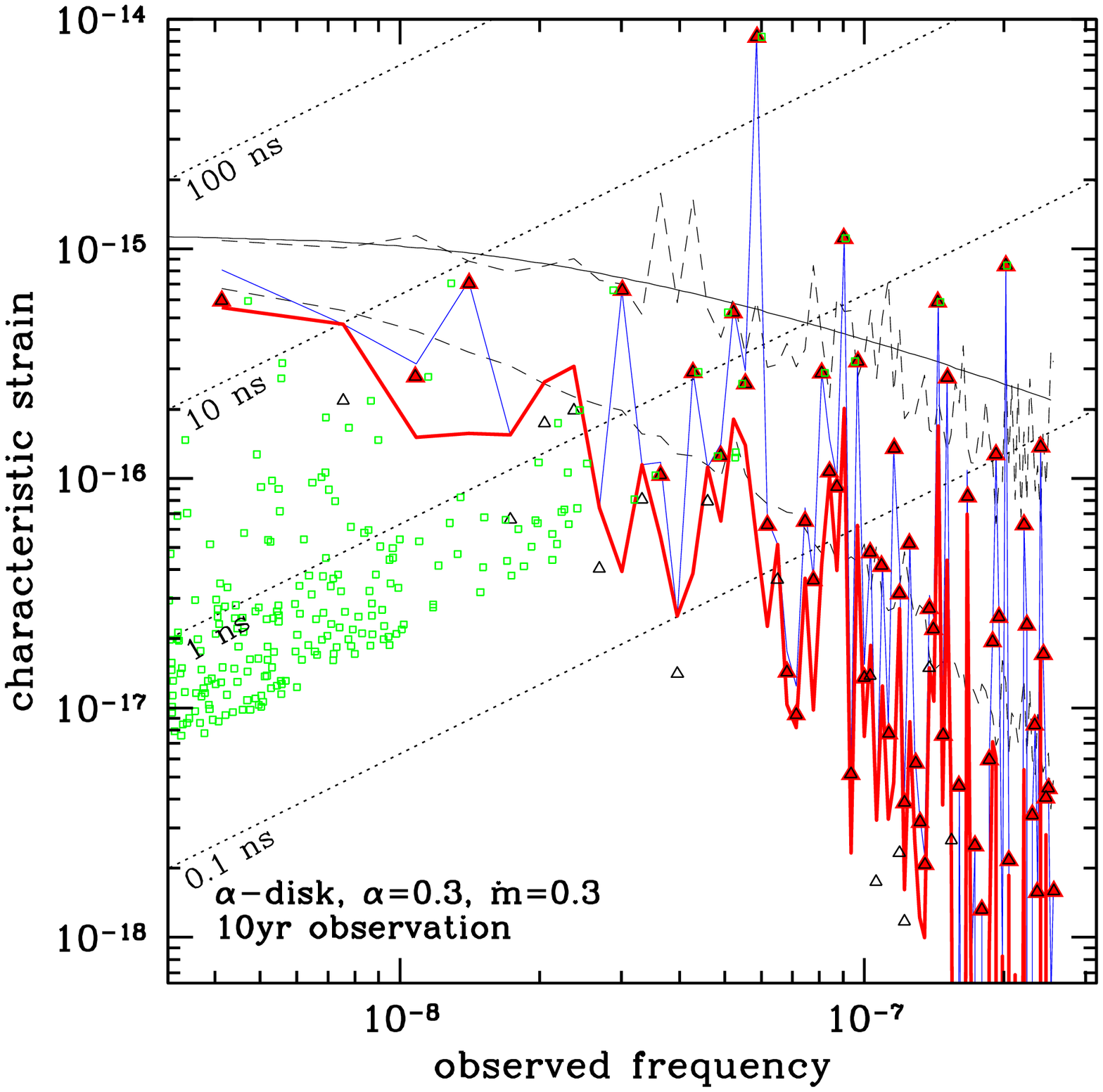}
   \includegraphics[clip=true]{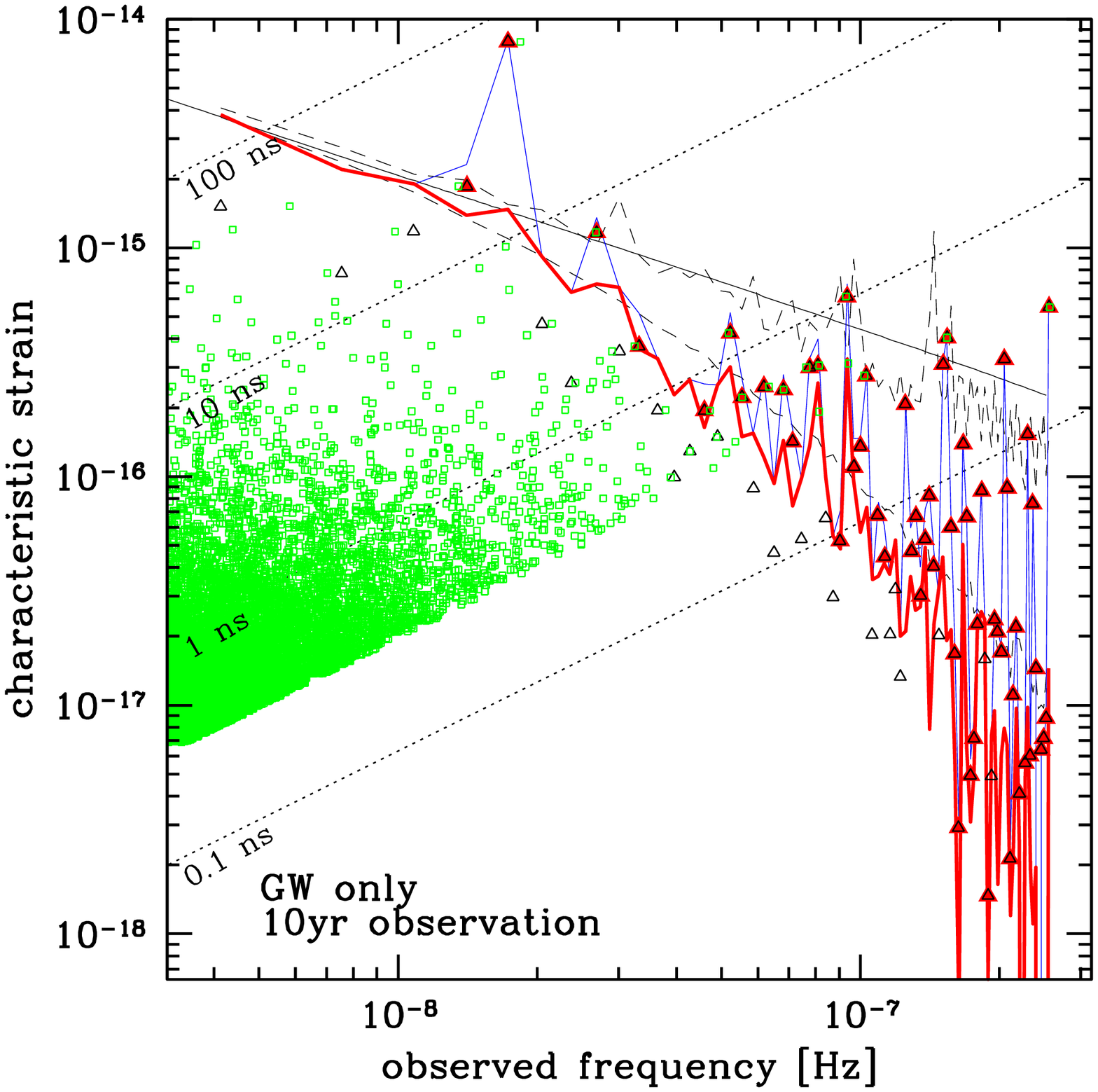}}
     \caption{{Components} of the GW signal from a population of inspiralling MBHBs. In the left panel we
consider {all} binaries embedded in a gaseous $\alpha$-disk for our default model (all mergers wet), while
in the right panel {all} binaries are purely GW driven (all mergers dry). In each panel, the smooth
solid line is the RMS total characteristic GW strain $h_c$ using the integral expression
(\ref{e:hc_GWdrivenN}--\ref{e:hc_GWdrivenN2}) (which correspond to an average over $N_k\rightarrow\infty$
Monte Carlo realizations), the two dashed black lines represent the RMS total signal (upper) and the RMS
stochastic background level (lower) averaged over $N_k=1000$ Monte Carlo realizations respectively. The jagged
blue line displays a random Monte Carlo realization of the GW signal. The small black and red triangles show the
contributions of the brightest and resolvable sources in each frequency bin respectively. The jagged red line is the stochastic
GW background for this realization, i.e., once the resolvable sources in each frequency bin are subtracted. The
green dots label all the systems producing an RMS residual $t_{\rm gw}>0.3\,$ns over $T=10$\,years. The dotted
diagonal lines shows constant $t_{\rm gw}$ levels as a function of frequency. An observation time of 10 years is assumed.
}
        \label{f1}
    \end{figure*}

As stated in Sec. 2.3, the relevant frequency band for pulsar
timing observations, assuming a temporal observation baseline $T$ and a time interval between subsequent observations
$\Delta t$, is between $f_{\rm min}\approx 1/T$ and $f_{\rm max}\approx 1/(2\Delta t)$
with a resolution $\Delta{f}\approx 1/T$. In our calculations we assume a default duration of $T=10$\,yr for the
PTA campaign with $\Delta t\approx 1$ week. This gives $f_{\rm min}\approx 3\times 10^{-9}$\,Hz, $f_{\rm max}\approx 10^{-6}$Hz
and $\Delta{f}\approx 1/T \approx 3\times 10^{-9}\,$Hz.
The simulated signal is computed by doing a Monte Carlo sampling
of the distribution $\partial^3N/(\partial z\partial \Mc \partial \ln f_r)$, and adding the GW contribution of
each individual source. In each frequency bin, we identify the individually resolvable sources and the stochastic background.
We repeat this exercise $N_k=1000$ times for the 12 steady disk models defined in Sec.~\ref{s:gasdetails} and for the purely
GW--driven case. In addition, we also calculate the GW signal using the integral expression (\ref{e:hc_GWdrivenN}--\ref{e:hc_GWdrivenN2})
using the continuous distribution function, which corresponds to the RMS average of the GW signal in the $N_k\rightarrow\infty$ limit.

The GW signal and its most important ingredients are plotted in Figure~\ref{f1}. The left panel shows
results for gas--driven inspirals using our default $\alpha$-disk (i.e. ``gas on'' -- all mergers wet), the right panel shows
results for purely GW--driven inspirals for comparison (i.e. ``gas off'' -- all mergers dry) for the same underlying cosmological
MBHB coalescence rate. A randomly selected Monte Carlo realization of the signal is depicted as a dotted blue jagged line.
The green dots represent the contributions of individual binaries; systems producing
$\delta t_{\gw}(f)>0.3\,$ns timing residual are shown.
In each frequency bin, the brightest source is marked by a black triangle. The individually resolvable
sources are marked by superposed red triangles, and the stochastic level from all unresolvable sources is
shown with a solid red jagged line. Clearly, the GW signal for any single realization is far
from being smooth. The noisy  nature of the signal is due to rare massive binaries rising well above
the stochastic level.
Solid black curves in Figure~\ref{f1} show $h_c(f)$, the RMS value of the GW signal averaged over the merger
episodes, using the integral expression (\ref{e:hc_GWdrivenN}--\ref{e:hc_GWdrivenN2}). The upper black dashed
curve is the GW signal, averaged over $N_k=1000$ Monte Carlo realizations of the merging systems. This is
still noisy, due to the finite number of realizations used, but is consistent with the integrated average
shown by the solid black curve. The lower black dashed curve marks the RMS stochastic component,
$h_{s,{\rm rms}}(f)$, averaged over the same $N_k=1000$ realizations.
Figure~\ref{f1} shows that the gaseous disk greatly reduces the number of binaries at small frequencies compared
to the GW--only case, as gas drives the binaries in quickly towards the final coalescence. Consequently, the signal is more spiky
in the presence of gas. There is a clear flattening of the RMS spectrum at low frequencies ($f<1/$yr) in the
gas-driven case compared to the purely GW-driven case. The stochastic level is more suppressed in the wet-only
case and is much steeper than $f^{-2/3}$ at high frequencies. Despite the spiky signal in a given Monte Carlo
realization, the overall shape of the background is well recognizable in the GW driven model, while the characterization
of the global shape of the signal in the gas driven case appears to be less viable.

Gas driven migration becomes more and more prominent at large binary separations, corresponding to large orbital times
or small GW frequencies. Therefore, to check the ultimate maximum impact of gas effects on future PTA detections, we
simulated the spectrum for a hypothetical $T = 100$\,yr observation baseline. Figure \ref{f100yr} shows a realization
of the spectrum for such an extended observation, assuming our default gas model (i.e. all mergers wet).
Given the extended temporal baseline, the minimum observable frequency is pushed down to $\sim3\times10^{-10}\Hz$,
where the spectrum of gas driven mergers is considerably flatter. Moreover, the
frequency resolution bin is then narrower, the number of sources per bin is much smaller, and more sources
become resolvable. At the smallest frequencies, the induced timing residual can be higher
than 1$\mu$s and more than 50 sources will be individually resolvable at 1ns precision level, some of them with an SNR as high as 100.
These numbers are not severely modified if considering purely GW--driven dry mergers only (see Fig.~\ref{f:N_resolvable(t_gw)} below).
{Interestingly, due to the high frequency resolution of such an extended observation,
the GW frequency of some of the resolvable binaries may evolve significantly during the observation
(e.g. a binary with masses $m_1=m_2=5\times 10^8\msun$ or higher are nonstationary
relative to the bin size at frequencies above $f\gsim 40\,$nHz, see equation~[\ref{e:fevolution}]). Therefore it may be possible
to detect the frequency evolution of individually resolvable binaries during such an extended monitoring campaign. }
However, this
computation is idealized: it is questionable if millisecond pulsars can maintain a ns timing stability over such
a long timescale, nonetheless, it points out the enormous capabilities of long term PTA campaigns.

\begin{figure}
\centering
\mbox{\includegraphics[width=84mm]{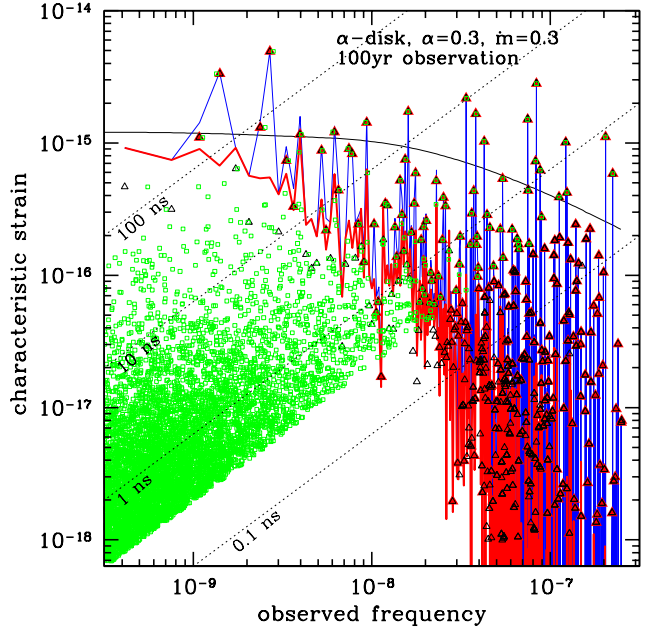}}
\caption{Same as left panel of Figure~\ref{f1}, but for an observation lasting 100 years (Averages over the 1000
realizations are not shown in this case for clarity).}

\label{f100yr}
\end{figure}

\subsection{Individually resolvable sources}

Let us now examine the prospects for detecting individual sources using PTAs.
How many sources are expected to be individually resolvable? How significant is their detection?

As explained in Sec.~\ref{s:timing}, the
signal from a specific binary can be detected using a PTA
if the corresponding timing residual $\delta t_{\gw}^2$ is above the RMS noise level $\delta t^2_{\rm N, rms}$ characterizing the PTA
given by Eq.~(\ref{e:snr}).
We should notice, however, that this estimate of the number of individually resolvable sources
is conservative and is likely to provide only a {\it lower limit} for the following reasons.
Firstly, we average over the sky position of the binaries and pulsars, while in reality, it may be possible
to take advantage of the different GW polarization and amplitude generated by sources in different sky positions to deconvolve their
signal (even if they have similar strength and frequency). Secondly, the brightest source identification algorithm can
be implemented recursively, after an accurate subtraction of the identified sources. However, we find that, especially
at low frequencies, the distribution of GW source amplitudes for various binaries in a single frequency bin is not
strongly hierarchical, so that a recursive brightest source finding algorithm shouldn't
increase significantly the number of resolvable systems.
We present here results both in terms of the total number ($N_t$) and resolvable systems ($N_r$, see Eq.~\ref{e:N}).

\begin{figure*}
\centering
\mbox{\includegraphics[width=144mm]{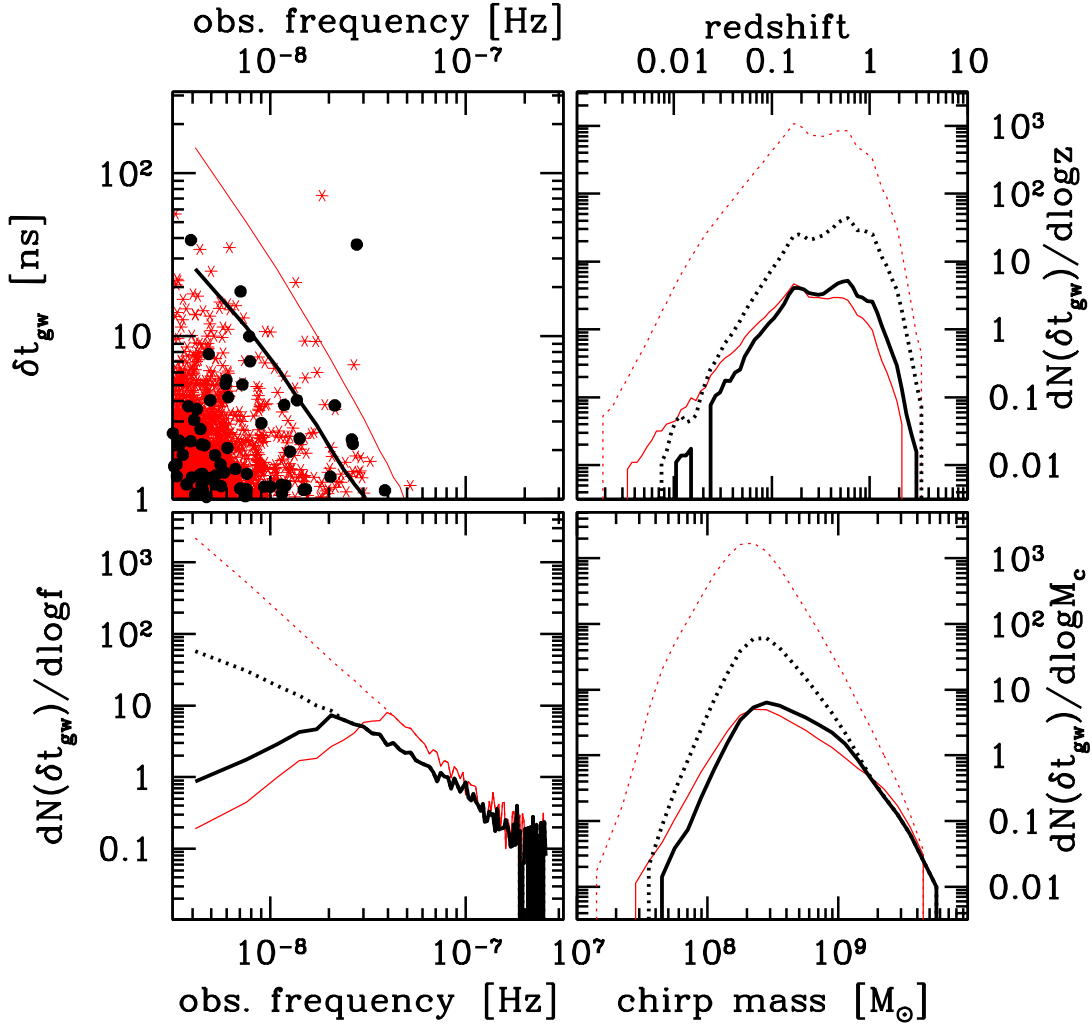}}
\caption{{\it Top-left panel}: characteristic amplitude of the
timing residuals $\delta t_\mathrm{gw}$ (equation (\ref{e:deltatgw})) as a function of frequency; the dots
are the residuals generated by individual sources and the solid line is the estimated {\it stochastic level}
of the GW signal. {\it Top-right panel}:
distribution of the number of total (dotted lines) and resolvable (solid lines) sources  per logarithmic
redshift interval as a function of redshift, generating a
 $\delta t_\mathrm{gw}>1$ns.
{\it Bottom-left panel}:
distribution of the number of total (dotted lines) and resolvable (solid lines) sources  per logarithmic
frequency interval as a function of the GW frequency, generating a
 $\delta t_\mathrm{gw}>1$ns.
{\it Bottom-right panel}: distribution of the total (dotted lines) and individually resolvable
(solid lines) number of sources per logarithmic chirp mass interval
as a function of chirp mass, generating a
 $\delta t_\mathrm{gw}>1$ns. All the black elements refer to our default disk model,
the red elements are for a GW driven MBHB population. Distributions are averaged over $N_k=1000$ realizations
of the MBHB population.
}
\label{f:N(t_gw)distribution}
\end{figure*}
In Figure~\ref{f:N(t_gw)distribution} we plot the distribution of the number of sources (total and resolvable) as a function of
timing residual, detection frequency, redshift, and chirp mass,
found in two particular realizations of our default $\alpha$-disk (all mergers wet) and in the purely GW-driven models (all mergers dry).
The figure shows that even though there are much more sources in the purely GW-driven case, the number of sources rising above
the stochastic level is almost the same in the purely dry and wet cases. Figure \ref{f:N(t_gw)distribution}
also shows the chirp mass, redshift, and frequency distribution for sources above 1\,ns timing level.
As was previously shown in SVV09, the bulk of the sources
are cosmologically nearby ($z\lsim 1$) with masses peaking around ${\cal M}\sim 2\times 10^8\msun$.
Figure \ref{f:N(t_gw)distribution} shows that gas dynamics does not introduce
a major systematic change in the shape of the redshift and the chirp mass distributions.
The lower left panel shows that gas removes a systematically larger fraction of sources at
small frequencies.

\begin{figure}
\centering
\mbox{\includegraphics[width=84mm]{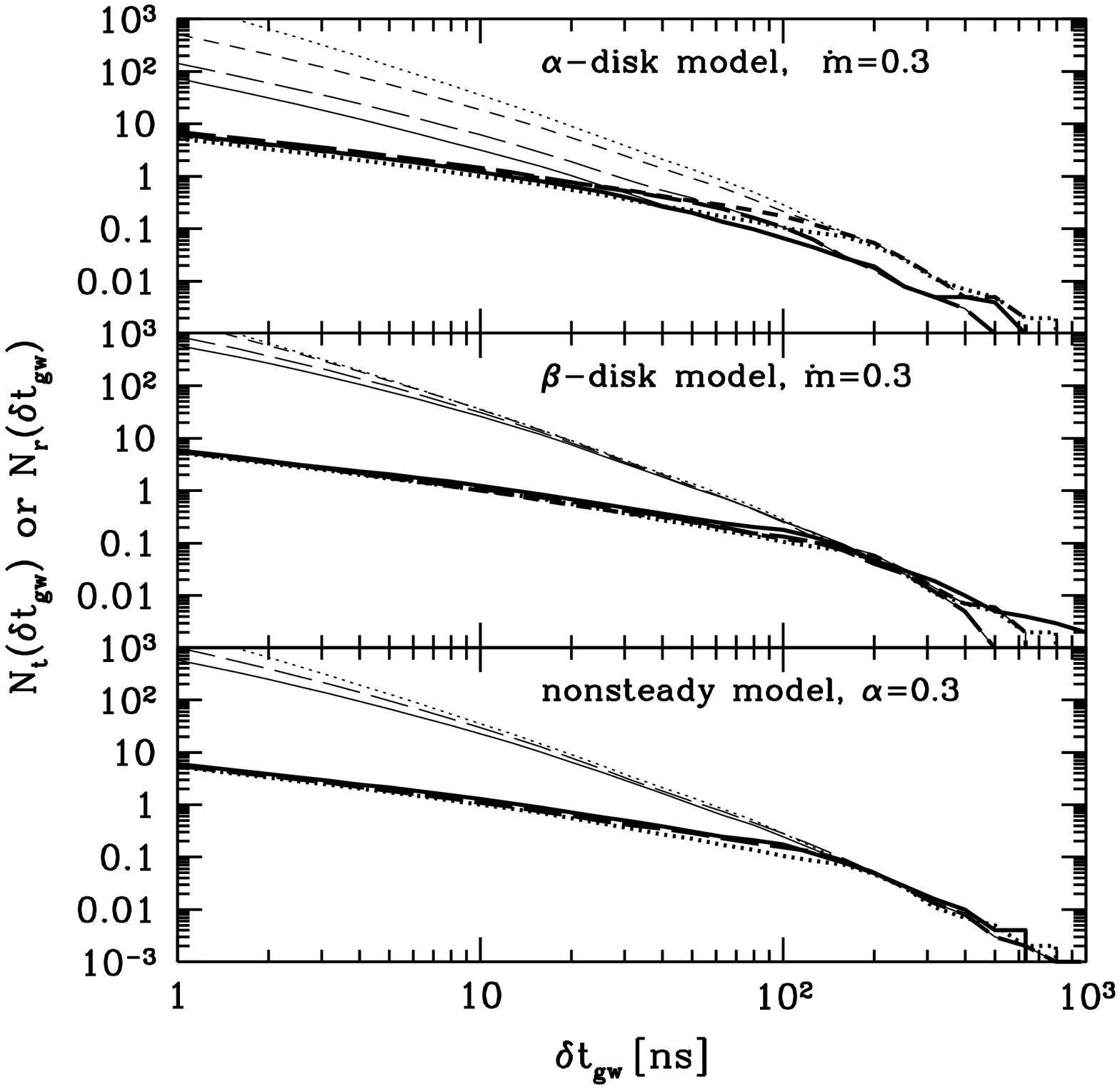}}
\caption{Cumulative number of total ($N_t(\delta t_{\rm gw})$, thin lines) and individually resolvable
($N_r(\delta t_{\rm gw})$, thick lines) sources emitting above a
given $\delta {t_{\rm gw}}$ threshold as a function of
$\delta {t_{\rm gw}}$. {\it Upper panel}: $\alpha$--disk model with $\dot{m}=0.3$.
Lines are for $\alpha=$0.3 (solid), 0.1 (long--dashed) and 0.01 (short--dashed).
{\it Central panel}: same as upper panel but for a $\beta$-disk model.
{\it Lower panel}: nonsteady IPP model, assuming $\alpha=0.3$ and $\dot{m}=0.3$
(solid) and 0.1 (long--dashed). In all the panels, the dotted lines refer, 
for comparison, to the GW driven model.
All the distributions refer to the ensemble mean computed over all $N_k=1000$
realizations of the MBHB population.}
\label{f:N_resolvable(t_gw)}
\end{figure}
\begin{figure}
\centering
\mbox{\includegraphics[width=84mm]{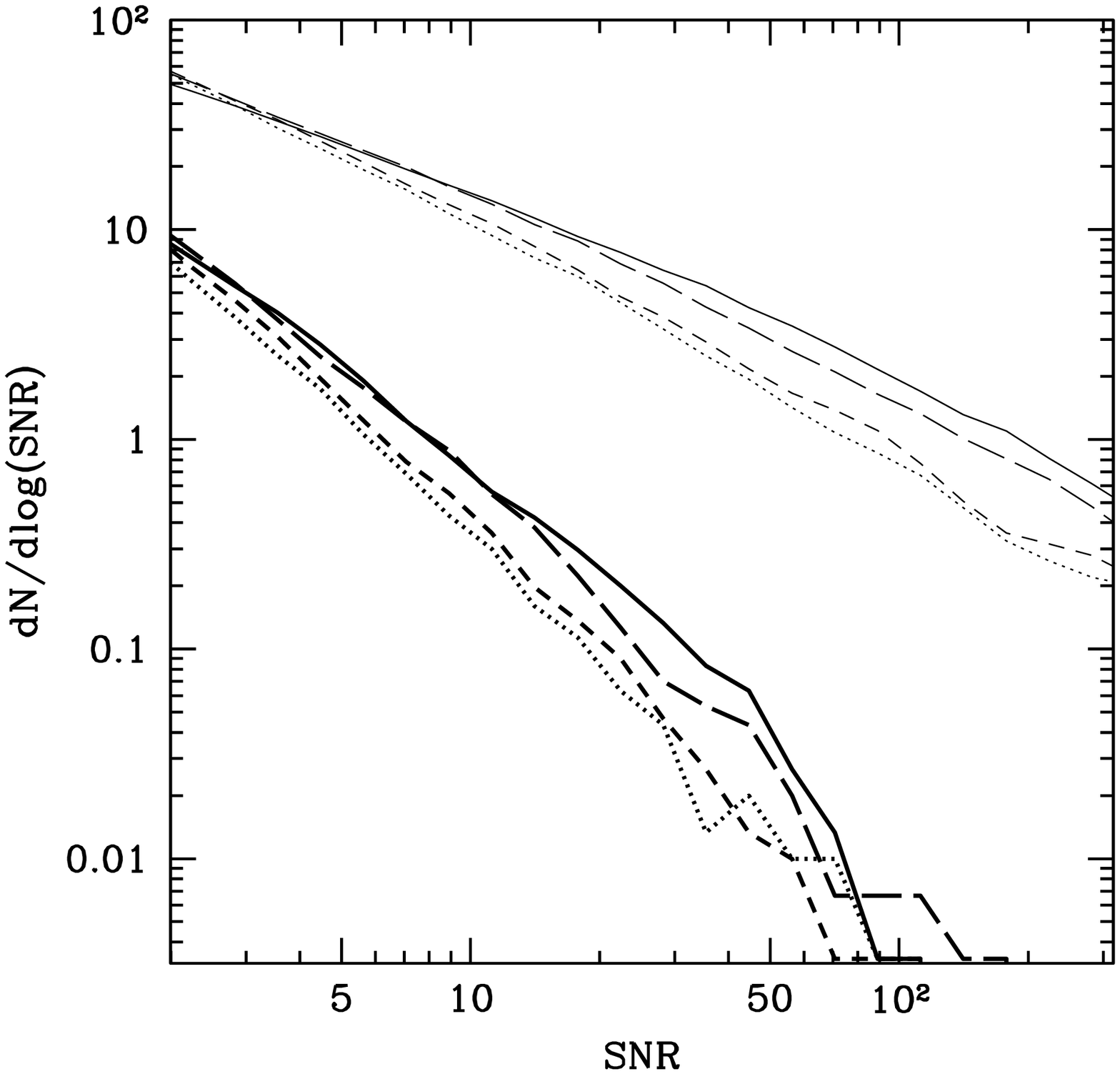}}
\caption{SNR distribution of individually resolvable sources. Thin and thick curves refer
to neglecting the instrumental noise or using a $\delta t_{\gw}=1\,$ns timing precision, respectively.
Linestyle as in the upper panel of Figure~\ref{f:N_resolvable(t_gw)}.}
\label{f6}
\end{figure}
Figure~\ref{f:N_resolvable(t_gw)} shows the cumulative number of binaries (total and resolvable) as a function of
timing residual. The upper panel shows results for the $\alpha$--disk
models with $\dot{m}=0.3$ and different $\alpha$. The statistics of resolvable
sources is almost unaffected by the large suppression of the total number of sources at a fixed timing residual. For example,
in all of our models, we expect $\sim 2$ resolvable sources at a timing level of $\delta t_{\gw} = 10$\,ns, even though the total number of
sources contributing to the signal at that level spans about an order of magnitude among the different models
($\sim 5$ for $\alpha=0.3$ to $\sim 50$ for binaries driven by GW only). The same is true for $\beta$--disk models
 { (central panel) and for nonsteady models (lower panel)},
even though in these cases the total number of sources at a particular $\delta t_{\gw}$ is not reduced dramatically by gas effects.
This result can be understood with a closer inspection of Figure~\ref{figtres}. Let us focus on the $\alpha$--disk model.
As explained in Sec. 3.2, the impact of gas driven dynamics in the PTA window is more significant for lower binary masses
and unequal mass ratios. These
are the binaries that build-up the bulk of the signal, and its stochastic level is consequently
greatly reduced in the gas driven case. On the other hand, the population of high equal mass binaries, which constitute most of the
individually resolvable sources,
 is almost unaffected
by the presence of the circumbinary disk, as they are already in the GW-driven regime in the relevant range of binary periods.

Figure~\ref{f6} shows expectations on the detection significance of resolvable sources. The
SNR distribution of resolvable sources (equation
[\ref{e:snr}]) are shown as thin lines, accounting for the astrophysical GW noise from the unresolved binaries,
but neglecting the intrinsic noise of the array (i.e. assuming an ideal detector with infinite
sensitivity, $\delta t_\mathrm{p}^2(f)=0$, in equation \ref{e:noise}). This calculation represents
{\it an upper limit} of the SNR. Thick lines, in Figure~\ref{f6} plot
the SNR considering a total detector noise of 1ns (appropriate for SKA).
The figure shows that the expected detection significance of resolvable sources is systematical higher for gas driven models with larger $\alpha$.
In general, in an array with 1ns sensitivity, we may expect a couple of individually resolvable sources with $\rm SNR>5$.
For near future instruments with much worse sensitivity, the identification of resolvable sources might be
more challenging.

\subsection{Stochastic background}
\begin{figure}
\centering
\mbox{\includegraphics[width=84mm]{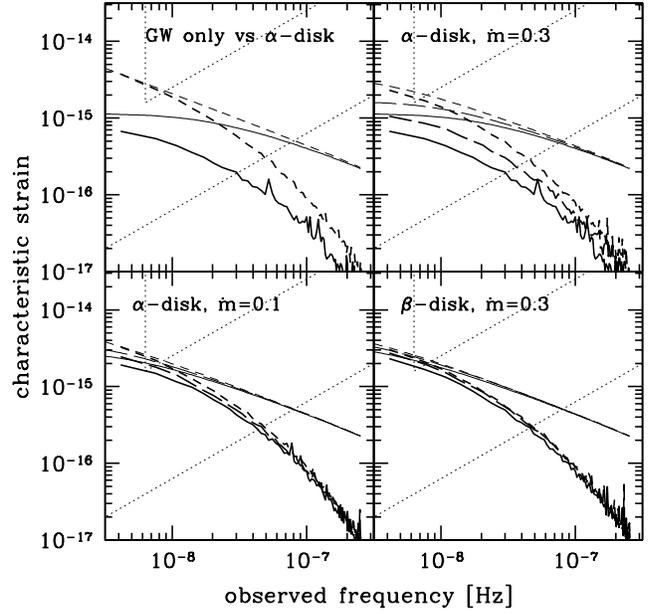}}
\caption{Influence of the gas driven dynamics on $h_c$ (thin lines) and on
 $h_s$ (thick lines). {\it Upper left panel}: GW driven dynamics versus gas driven dynamics for our default disk model.
{\it Upper right panel}: $\alpha=$ 0.3 (solid), 0.1 (long--dashed),
0.01 (short--dashed), for an $\alpha$--disk with $\dot{m}=0.3$.
{\it Lower left panel}: same as the upper right panel, considering an $\alpha$--disk with $\dot{m}=0.1$.
{\it Lower right panel}: same as the upper right panel, considering a $\beta$--disk with $\dot{m}=0.3$.
The two dotted lines in each panel represent the sensitivity of the complete PPTA survey and an indicative
sensitivity of 1ns for the SKA.
}
\label{f:stochasticRMS}
\end{figure}

Figure \ref{f:stochasticRMS} shows the RMS stochastic level (i.e. after subtracting off the individually resolvable sources,
see Sec.~\ref{s:statistics}) for selected steady state gas disk models, averaged over $N_k=1000$ realizations.
The top left panel highlights the difference between wet and dry models, the top right and bottom left panels collect different
$\alpha$--disk models and the lower right panel is for selected $\beta$--disk models 
{(nonsteady models, not shown here, give results basically identical to $\beta$--disk models)}.
The different line styles show the effect of changing the $\alpha$ parameter of the disk.
We also show the RMS total signal level, which is exactly proportional to $f^{-2/3}$ in the dry case. In general,
the stochastic level matches the total signal level at low frequencies, but is increasingly suppressed for
frequencies above $f\gsim 10^{-8}\Hz$ for all of our models.
At sufficiently large frequencies GW emission dominates even for wet mergers, and
both the RMS total signal and the stochastic level approaches the purely GW--driven case.
However, at small frequencies, a significant fraction of binaries is driven by gas,
and the signal is attenuated and the spectrum is less steep
compared to the dry case. Figure \ref{f:stochasticRMS} shows that gas--driven
migration suppresses the stochastic background significantly, by a factor of 5 for
our standard disk model below $10^{-8}\Hz$.
The suppression of the total and stochastic levels is a strong function of the model parameters.
Interestingly, there is almost no suppression for $\beta$--disks, or for $\alpha$--disks with a
small accretion rate and/or a small $\alpha$ value. In these cases, the local disk mass is smaller,
resulting in longer viscous timescales, and hence the population of widely separated binaries
is not reduced significantly.

\begin{figure}
\centering
\mbox{\includegraphics[width=84mm]{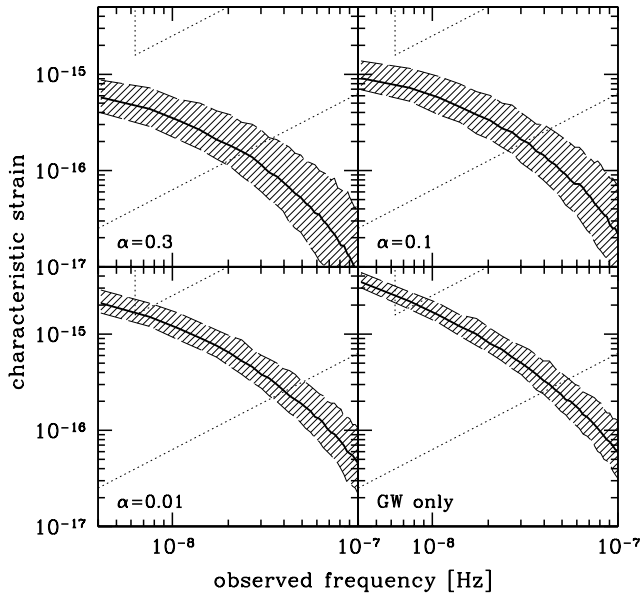}}
\caption{Variance of the expected stochastic level of the signal as a function of $\alpha$. In all
the panels we assume the $\alpha$--disk model with $\dot{m}=0.3$ (except for the lower right panel,
referring to the GW driven model). Solid lines represent the median $h_s$ over 1000 Monte Carlo realizations,
while the shaded area enclosed within the two dashed lines is the $10\%$--$90\%$ confidence region . The two dotted
lines in each panel represent the sensitivity of the complete PPTA survey and an indicative sensitivity of 1ns for the SKA.
}
\label{f:stochasticVAR}
\end{figure}
Figure~\ref{f:stochasticVAR} quantifies how the stochastic background changes among different
Monte Carlo realizations, showing the range attained by 10--$90\%$ of all $N_k=1000$ realizations.
The variance is not the product of uncertainties related to the parameters of
the adopted cosmological and dynamical model, but is purely determined by the small number statistics
of sources per frequency bin, intrinsic to
the source distribution.  At $f=10^{-8}$Hz, the variance of the signal produced by our default disk model
is $\sim 0.5$dex. Decreasing $\alpha$ to 0.1 and 0.01, the variance drops to $\sim0.35$dex and $\sim0.25$dex
respectively. In the case of all mergers driven by GWs, the variance at the same frequency is only $\sim0.15$dex.
This means that, contrary to the GW driven model (SVC08), it is impossible to predict the stochastic level of the signal
accurately for our default gas driven model. For a linear fit, any power law in the range $f^{-0.3}-f^{-1.1}$
is acceptable within the level of variance of the signal in the frequency range $3-30$\,nHz .

\section{Discussion and conclusions}
The presence of a strong nHz gravitational wave signal from a cosmological population of massive
black hole binaries, is a clear prediction of hierarchical models of structure formation, where
galaxy evolution proceeds through a sequence of merger events. The detailed nature of the signal depends
however on a number of uncertain factors: the MBHB mass function, the cosmological merger rate,
the detailed evolution of binaries,
and so on. In particular, MBHB dynamics determines the number of
sources emitting at any given frequency, and thus the overall shape and strength of the signal. Previous works
on the subject (e.g. SVC08 and SVV09) considered the case of GW driven binaries only.
In this paper we studied the impact of gas driven massive black hole binary dynamics on the nHz
gravitational wave signal detectable with pulsar timing arrays. This is relevant because in any
merger event, cold gas is efficiently funnelled toward the centre of the merger remnant, providing a
large supply of gas to the MBHB formed following the galaxy interaction.
Even a percent of the galaxy mass in cold gas funnelled toward the centre, is much larger than the masses
of the putative MBHs involved in the merger so that the newly--formed binary evolution may be driven
by gas until few thousand years before final coalescence.

To conduct our study, we coupled
models for gas driven inspirals of HKM09 to the MBHB population models presented
in SVV09. Our simulations cover a large variety of steady--state and quasistationary one-zone disk models for the
MBHB-disk dynamical interactions (with an extensive exploration of the $\alpha$ viscosity parameter and $\dot{m}$
for $\alpha$ and $\beta$-disks), along with several different prescriptions
for the merging MBHB population (four different black hole mass--galaxy bulge relations coupled with
three different accretion recipes).
The differences with respect to the purely GW driven models (presented
in SVV09) are qualitatively similar for all the considered MBHB populations, we thus presented the results
for the Tu-SA population model only, focusing on the impact of the different disk models.
Our main findings can be summarized as follows:
\begin{itemize}
\item The effect of gas driven dynamics may or may not be important depending on the properties of the
circumbinary disk. A robust result is that if the viscosity is proportional to the gas
pressure only ($\beta$-disk models), there is basically no effect on the GW signal, independently of the other disk parameters.
{ Similarly, there is a tiny effect for time-dependent migration models \citep[i.e.][]{ivanov}, where gas piles up and the accretion rate decreases as the binary hardens.
However, if the viscosity is proportional to the total (gas+radiation) pressure \citep[$\alpha$-disk,][]{ss73},
then the GW signal can be significantly affected for certain steady state circumbinary disk models 
\citep{syerclarke} with $\alpha\gsim 0.1$ and $\dot{m}\gsim 0.3$.}
This difference is explained by the fact that the gas--driven inspiral rate is determined by the radial gas inflow velocity, which is significantly faster for radiation pressure dominated $\alpha$-disks for a fixed accretion rate, as they are more dilute. Therefore gas effects can dominate over the GW--driven inspiral rate for $\alpha$ disks at relatively small binary separations corresponding to the PTA frequency band.
\item With respect to the GW driven case, the presence of massive circumbinary disks affects the
population of low--unequal mass binaries predominantly ($M<10^8\msun$, $q<0.1$), causing a significant suppression of the
{\it stochastic level} of the signal, but leaving the number and strength of massive {\it individually
resolvable sources} basically unaffected. In our default model ($\alpha=0.3$, $\dot{m}=0.3$), the stochastic background
is suppressed by a factor of $\sim 5$ at $f<10^{-8}$Hz. This suppression factor {decreases for smaller $\dot{m}$
and $\alpha$. The stochastic level is {\it not} suppressed significantly for nonsteady disks, for $\beta$--disks (arbitrary $\dot{m}$ and $\alpha$), and for $\alpha$-disks with either $\dot{m}\lsim 0.1$ and arbitrary $\alpha$, or $\alpha\lsim 0.01$ with arbitrary $\dot{m}$.}
{\it About 10 individual sources are resolvable at 1\,ns timing level, independently of the
adopted disk model.}
\item All the results shown here for the Tu-SA model, hold for every other MBHB population model we tested.
There is a certain level of degeneracy between disk dynamics and MBHB mass function: the stochastic level
given by a population of heavy binaries evolving by gas dynamics, can mimic that of
a population of lighter binaries that are driven by GWs only. However, the variance
of the signal would be much bigger in the former case, because the signal is produced by fewer massive sources.
\item The detection of GWs emitted by MBHBs embedded in gaseous disks with high viscosity and accretion rate,
may be very challenging for relatively short term PTA campaigns like the PPTA. In fact, we find that most of the 12 MBHB population
models tested in SVV09 would {\it not} produce a stochastic signal detectable by the PPTA (i.e. the signal is
a factor of three below the PPTA capabilities for the Tu-SA model). However, long term projects like the
SKA, which aim to nanosecond sensitivities, are expected to be able to detect the GW signal, resolving a handful of
individual sources with high significance.
\end{itemize}

A word of caution should be spent to stress the fact that our models are idealized in many ways. We considered
circular binaries only. If most systems were significantly eccentric, then the overall signal would be modified
by the multi-harmonic emission of each individual source.
Moreover, we only considered radiatively efficient, geometrically thin, one-zone, { steady--state and quasistationary 
accretion disks. 
The most massive but gas driven binaries, emitting at low GW frequencies $f_{\rm gw}\lsim 5\,$nHz, are 
radiation pressure dominated, but the migration rate estimates had to be extrapolated using the 
scaling exponents for gas pressure dominated disks. 
For even wider massive binaries ($f_{\rm gw}$ near the low frequency observation limit) 
the disks are marginally Toomre unstable.
We used simple models for the binary disk interaction, scaling the Type-II planetary migration formulae to
MBHBs. The results are sensitive to the models: the steady--state models of \citet{syerclarke} lead to a 
decrease in the stochastic nHz background, while the more sophisticated time-dependent models of 
\citet{ivanov} as well as the \citep{Hayasaki09,Hayasaki10} models
lead to almost no effect (see also Sec.~\ref{s:gasdetails}, for a list of caveats). }
Further studies should examine the accuracy of these approximations for gas driven migration in circumbinary disks around MBHBs.
Finally, it is likely that not all the binaries are
gas driven on their way to the coalescence, and the efficiency of the disk-binary coupling may vary from merger to
merger depending on the environmental conditions, which may modify the properties of the expected signal as well.
Nonetheless, our calculations provide clear predictions for the possible attenuation of the stochastic GW background,
which may be confirmed or discarded by ongoing and forthcoming pulsar timing arrays.

\section*{Acknowledgments}
A.S. is grateful to M. `penguin' Giustini and V. Cappa for enlightening discussions.
B.K. acknowledges support by NASA through Einstein Postdoctoral Fellowship grant number PF9-00063
awarded by the Chandra X-ray Center, which is operated by the Smithsonian Astrophysical Observatory for
NASA under contract NAS8-03060, and partial support by OTKA grant 68228.


\begin{thebibliography}{}

\bibitem[Amaro-Seoane et al. (2010)]{pau09} Amaro-Seoane P., Sesana A., Hoffman L., Benacquista M., Eichhorn C., Makino J., Spurzem R., 2010, MNRAS, 402, 2308

\bibitem[Artymowicz \& Lubow(1994)]{al94} Artymowicz P.,  Lubow S. H., 1994, ApJ, 421, 651

\bibitem[Artymowicz \& Lubow(1996)]{al96} Artymowicz P., Lubow S. H. 1996, ApJ, 467, L77

\bibitem[Armitage \& Natarajan (2002)]{an02} Armitage P. J., Natarajan, P., 2002, ApJ, 567, L9

\bibitem[Armitage \& Natarajan (2005)]{an05} Armitage P. J., Natarajan P., 2005, ApJ, 634, 921

\bibitem[Barnes (2002)]{barnes} Barnes J. E., 2002, MNRAS, 333, 481

\bibitem[Begelman, Blandford, \& Rees(1980)]{BBR} Begelman M. C.,  Blandford R. D.,   Rees M. J., 1980, Nature, 287, 307

\bibitem[Bender et al.(1998)]{Bender98} Bender P. et al., 1998, {\it LISA}, Laser Interferometer Space Antenna
for gravitational wave measurements: ESA Assessment Study Report

\bibitem[Bertone et al.(2007)]{Bertone07} Bertone S., De Lucia G.,  Thomas P. A., 2007, MNRAS, 379, 1143

\bibitem[Bertotti et al.(1983)]{Bertotti83} Bertotti B., Carr B. J.,  Rees M. J., 1983, MNRAS, 203, 945

\bibitem[Callegari et al.(2009)]{cal09} Callegari S., Mayer L., Kazantzidis S., Colpi M., Governato F., Quinn T.,  Wadsley J., 2009, ApJ, 696, 89

\bibitem[Cuadra et al.(2009)]{cua09} Cuadra J., Armitage P. J., Alexander R. D.,  Begelman M. C., 2009, MNRAS, 393, 1423

\bibitem[Detweiler (1979)]{Detweiler79} Detweiler S., 1979, ApJ, 234, 1100

\bibitem[Dotti et al.(2007)]{dotti07} Dotti M., Colpi M., Haardt F.,  Mayer, L., 2007, MNRAS, 379, 956

\bibitem[Dubus, Hameury, \& Lasota(2001)]{Dubus01} Dubus G., Hameury J.-M., Lasota J.-P., 2001, A\&A, 373, 251

\bibitem[Edwards, Hobbs, \& Manchester (2006)]{ed06} Edwards R. T., Hobbs G. B.,  Manchester R. N., 2006, MNRAS, 372, 1549

\bibitem[Escala et al.(2004)]{escala04} Escala A., Larson R. B., Coppi P. S.,  Mardones D., 2004, ApJ, 607, 765

\bibitem[Escala et al.(2005)]{escala05} Escala A., Larson R. B., Coppi P. S.,  Mardones D., 2005, ApJ, 630, 152

{ \bibitem[Goldreich \& Tremaine(1979)]{gt79} Goldreich P., Tremaine S., 1979, ApJ, 233, 857}

\bibitem[Haehnelt (1994)]{Haehnelt94} Haehnelt M. G., 1994, MNRAS, 269, 199

\bibitem[Haiman, Kocsis, \& Menou (2009);HKM09]{HKM09} Haiman Z., Kocsis B.,  Menou K., 2009, ApJ, 700, 1952 (HKM09)

{ \bibitem[Hayasaki (2009)]{Hayasaki09} Hayasaki K., 2009, PASJ, 61, 65}

{ \bibitem[Hayasaki et al.(2010)]{Hayasaki10} Hayasaki K., Ueda Y., Isobe N., 2010, eprint arXiv:1001.3612}

\bibitem[Hellings \& Downs (1983)]{Hellings83} Hellings R. W.,  Downs G. S., 1983, ApJ, 265, 39

\bibitem[Hirose, Krolik, \& Blaes (2009)]{hirose09} Hirose S., Krolik J. H.,  Blaes O., 2009, ApJ, 691, 16

\bibitem[Hobbs et al. (2010)]{Hobbs09} Hobbs G. et al., 2010, CQGra, 27, 4013

\bibitem[Ivanov, Papaloizou, \& Polnarev(1999)]{ivanov} Ivanov P. B., Papaloizou J. C. B.,  Polnarev A. G., 1999, MNRAS, 307, 79

\bibitem[Janssen et al.(2008)]{Janssen08} Janssen G. H., Stappers B. W., Kramer M., Purver M., Jessner A.,  Cognard I., 2008,
in ``40 YEARS OF PULSARS: Millisecond Pulsars, Magnetars and More'', AIP Conference Proceedings, 983, 633

\bibitem[Jaffe \& Backer (2003)]{Jaffe03} Jaffe A. H.,  Backer D. C., 2003, ApJ, 583, 616

\bibitem[Jenet et al. (2005)]{Jenet05} Jenet F. A., Hobbs G. B., Lee K. J.,  Manchester R. N., 2005, ApJ, 625, 123

\bibitem[Jenet et al. (2009)]{jen09} Jenet R. et al., 2009, arXiv:0909.1058

\bibitem[King, Pringle, \& Livio(2007)]{King07} King A. R., Pringle J. E.,  Livio M., 2007, MNRAS, 376, 1740

\bibitem[Kollmeier et al.(2006)]{Kollmeier06} Kollmeier J. A. et al., 2006, ApJ, 648, 128

\bibitem[Koushiappas \& Zentner (2006)]{Koushiappas06} Koushiappas S. M.,  Zentner A. R., 2006, ApJ, 639, 7
	
\bibitem[Kocsis, Haiman, \& Menou(2008)]{Kocsis08} Kocsis B., Haiman Z.,  Menou K., 2008, ApJ, 684, 870

\bibitem[Lazio (2009)]{Lazio09} Lazio J., 2009, arXiv:0910.0632

\bibitem[Lightman \& Eardley(1974)]{le74} Lightman A. P., Eardley D. M., 1974, ApJ, 187, L1

\bibitem[Lodato et al.(2009)]{lodato09} Lodato G., Nayakshin S., King A. R., Pringle J. E., 2009, MNRAS, 398, 1392

\bibitem[Lubow \& D'Angelo(2006)]{LubowDAngelo06} Lubow S. H., D’Angelo G. 2006, ApJ, 641, 526

\bibitem[Lubow et al.(1999)]{Lubow99} Lubow S. H., Seibert M., Artymowicz P., 1999, ApJ, 526, 1001

\bibitem[MacFadyen \& Milosavljevi\'c(2008)]{mm08} MacFadyen A., Milosavljevi\'c M., 2008, ApJ, 672, 83

\bibitem[Malbon et al.(2007)]{Malbon07} Malbon R. K., Baugh C. M., Frenk C. S.,  Lacey, C. G., 2007, MNRAS, 382, 1394

\bibitem[Manchester (2008)]{Manchester08} Manchester R. N., 2008, in ``40 YEARS OF PULSARS: Millisecond Pulsars, Magnetars and More'', AIP Conference Proceedings, 983, 584

\bibitem[Milosavljevic \& Merritt (2003)]{Milos03} Milosavljevic M.,  Merritt D., 2003, ApJ, 596, 860

{ \bibitem[Paardekooper \& Papaloizou (2009)]{Paardekooper1} Paardekooper S.-J., Papaloizou J.C.B., 2009, MNRAS, 394, 2283}

{ \bibitem[Paardekooper et al.(2010)]{Paardekooper2} Paardekooper S.-J., Baruteau C., Crida A., Kley W., 2010, MNRAS, 401, 1950}

\bibitem[Piran(1978)]{piran78} Piran T., 1978, ApJ, 221, 652

\bibitem[Pessah, Chan, \& Psaltis(2007)]{Pessah07} Pessah M. E., Chan C-K., Psaltis D., 2007, ApJ, 668, 51L

\bibitem[Phinney(2001)]{Phinney01} Phinney E. S., 2001, arXiv:astro-ph/0108028

\bibitem[Sazhin (1978)]{Sazhin78} Sazhin M. V., 1978, Soviet Astron., 22, 36

\bibitem[Sesana et al.(2004)]{Sesana04} Sesana A., Haardt F., Madau P.,  Volonteri, M., 2004, ApJ, 611, 623

\bibitem[Sesana et al.(2005)]{Sesana05} Sesana A., Haardt F., Madau P.,  Volonteri, M., 2005, ApJ, 623, 23

\bibitem[Sesana, Vecchio, \& Colacino (2008)]{SVC08} Sesana A., Vecchio A.,  Colacino C. N., 2008, MNRAS, 390, 192 (SVC08)

\bibitem[Sesana, Vecchio, \& Volonteri (2009)]{SVV09} Sesana A., Vecchio A.,  Volonteri M., 2009, MNRAS, 384, 2255 (SVV09)

\bibitem[Sesana \& Vecchio (2010)]{SV10} Sesana A., Vecchio A., 2010, Phys. Rev. D., 81, 4008

\bibitem[Shakura \& Sunyaev(1973)]{ss73} Shakura N. I.,  Syunyaev R. A., 1973, A\&A, 24, 337

\bibitem[Springel et al.(2005)]{springel} Springel V. et al., 2005, Nature, 435, 629

\bibitem[Syer \& Clarke(1995)]{syerclarke}	Syer D., Clarke C. J., 1995, MNRAS, 277, 758

\bibitem[Thorne (1987)]{Thorne87} Thorne K. S., 1987, in 300 Years of Gravitation, ed. S. Hawking \& W.
Israel (Cambridge: Cambridge Univ. Press), 330

\bibitem[Trump et al.(2009)]{Trump09} Trump J. R. et al., 2009, ApJ, 700, 49

\bibitem[Tundo et al. (2007)]{Tundo07} Tundo E., Bernardi M., Hyde J. B., Sheth R. K.,  Pizzella A., 2007, ApJ, 663, 57

\bibitem[Volonteri, Haardt, \& Madau(2003)]{VHM03} Volonteri M., Haardt F.,  Madau P., 2003, ApJ, 582, 599

{ \bibitem[Tanaka et al.(2002)]{ttw02} Tanaka H., Takeuchi T.,  Ward W. R., 2002, ApJ, 565, 1257}

\bibitem[Wyithe \& Loeb (2003)]{Wyithe03} Wyithe J.~S.~B., Loeb A., 2003, ApJ, 590, 691

\bibitem[Yoo et al. (2007)]{Yoo07} Yoo J., Miralda-Escud\'e J., Weinberg D.~H., Zheng Z.,  Morgan C.~W., 2007, ApJ, 667, 813

\end{thebibliography}
\end{document}